\PassOptionsToPackage{sort&compress}{natbib} % reference ranges

\documentclass[preprint,review,10pt,3p]{elsarticle}

\usepackage{amssymb}
\usepackage{amsmath}
\usepackage{lineno}

\usepackage[table]{xcolor}
\usepackage{lipsum}
\usepackage{hyperref}
\usepackage{xurl}
\usepackage{tabularx}
\usepackage{booktabs}
\usepackage{multicol}
\usepackage{multirow}
\usepackage{makecell}
\usepackage{subfig}
\usepackage{cleveref}
\usepackage{siunitx}
\usepackage{eurosym}
\usepackage{setspace}
\usepackage[bottom]{footmisc} % footnotes at bottom of page
\newcolumntype{Y}{>{\centering\arraybackslash}X} % centred column in tabularx

\usepackage{amsmath,amssymb,amsfonts} % for subsequations
\usepackage{relsize} % for mathlarger
\usepackage{gensymb}
\usepackage{commath}
\usepackage{mathrsfs}
\usepackage{bbm}
\usepackage{pifont}
\usepackage{array}

\makeatletter
\newcommand{\vast}{\bBigg@{3}}
\newcommand{\Vast}{\bBigg@{4}}
\makeatother

\newcommand{\edit}[1]{\textcolor{black}{#1}}
\newcommand{\edittwo}[1]{\textcolor{black}{#1}}

\journal{Energy}

\begin{document}

\begin{frontmatter}



\title{The value of hedging against energy storage uncertainties when designing energy parks}


\author[EECi]{Max Langtry\corref{cor1}} \ead{mal84@cam.ac.uk}
\author[EECi,ATI]{Ruchi Choudhary}

\cortext[cor1]{Corresponding author}

\affiliation[EECi]{organization={Energy Efficient Cities Initiative, Department of Engineering, University of Cambridge},
            addressline={Trumpington Street},
            city={Cambridge},
            postcode={CB2 1PZ},
            country={UK}}
\affiliation[ATI]{organization={Data-Centric Engineering, The Alan Turing Institute},
            addressline={British Library},
            city={London},
            postcode={NW1 2DB},
            country={UK}}

%% Abstract
\begin{abstract}
% 250 words
Energy storage is needed to match renewable generation to industrial loads in energy parks. However, the future performance of bulk storage technologies is currently highly uncertain. Due to the urgency of decarbonization targets, energy park projects must be designed and begun now. But, as uncertainty in storage performance reduces, a different technology than identified during initial design may turn out cheaper. Enabling flexibility so that designs can be updated as better information becomes available would lower the cost of decarbonizing industry. But having this flexibility is itself costly. This raises the question, ``Is it worth it?''

This study quantifies the benefit of retaining flexibility to adapt energy park designs and optionality over storage technology choice as uncertainty reduces, to determine whether it is economically worthwhile. It applies the Value of Information analysis framework to the sizing of wind, solar, and storage in an illustrative energy park model based on a real-world proposal near Rotterdam, considering uncertainty in storage efficiency, lifetime, and capital cost.

Updating asset sizings after storage uncertainty reduced is found to reduce total costs by 18\% on average. Having the option to switch storage technology choice as well reduces costs by a further 13\%, which is substantially greater than the cost of providing storage optionality. Using two storage technologies in the energy park reduces costs by 14\%, and in this case storage optionality is not worthwhile. These results are robust to the level of uncertainty reduction in storage performance, and the risk aversion of the system designer.
\end{abstract}

% \includegraphics[width=\linewidth]{figures/graph_abst.pdf}

\begin{highlights}
\item Sizing of wind, solar, and storage in energy park using stochastic programming
\item Impact of reducing uncertainty in storage efficiency, lifetime, and cost on design
\item Updating sizings after storage uncertainty reduced lowers cost by 18\% on average
\item Option to switch storage choice reduces costs by extra 13\%, more than cost of option
\item Using two storage techs reduces costs by 14\%, and optionality no longer worthwhile
\end{highlights}

%% Keywords

\end{frontmatter}

%% Add \usepackage{lineno} before \begin{document} and uncomment following line to enable line numbers
%\linenumbers

%% main text - 7000 word limit

\newpage
\section{Introduction} \label{sec:intro}

% aim for c. 1500 words
% for Energy example see https://www.sciencedirect.com/science/article/pii/S0360544221009403

%\subsection{Background}

% - electricity must be decarbonized to meet 2050 emissions goals
% - energy parks with storage are being considered as way to manage variability of wind generation and improve profitability
% - however future storage technology perform is highly uncertain, there is great disagreement in the literature
% - actual storage performance has significant impact on energy park cost, but is only revealed late in the design stage
% - while uncertainty can be accounted for during design, retaining optionality until after uncertainty revealed may be beneficial for energy park developers (and has not yet been studied)

Decarbonizing electricity generation is critical to achieving 2050 net-zero carbon emissions targets \cite{committeeonclimatechange2020SixthCarbonBudgetb}. Currently 40\% of electricity in the UK is generated from carbon intensive sources \cite{desnz2024DigestUKEnergy}, and electricity demand is expect to increase by a factor 3 to 4 as heating and transport energy usage are decarbonized via electrification \cite{nationalgrideso2023FutureEnergyScenarios}. As a result, around 100 GW of wind and 60 GW of solar generation need to be constructed by 2050 \cite{nationalgrideso2023FutureEnergyScenarios}.
However, managing the variability of electricity generation from these renewable sources to maintain security of supply presents a significant challenge \cite{papadis2020ChallengesDecarbonizationEnergy}, and the required adaptation of the energy system will be costly.
Grid-scale energy storage is expected to be the main mechanism for matching renewable electricity supply with demand. % As distributed storage is expensive, demand-side flexibility is limited in scope, and CCS-gas is un-realistic
Co-locating bulk energy storage with renewable generation and industrial loads, referred to as an energy park, has been proposed as a way of both improving the profitability of grid-scale storage (reducing the cost of its services), and reducing the grid impact of variable renewable generation \cite{chinaris2025HybridizationWindFarms}. %% + fan2021SizingCoordinationStrategies <- for thesis
% by keeping power fluctuations local, reducing the need for transmission line expansions

\edit{The design of energy parks, including the selection and sizing of renewable generation and energy storage technologies, has been extensively studied in the literature.
For example, Arévalo \& Jurado \cite{arevalo2021PerformanceAnalysisPV} investigates the combination of energy storage technology and control scheme that provides the lowest net present cost for supporting an autonomous grid with solar, wind, hydro, and diesel generation. Sizing optimization is performed for each system configuration, and it is found that due to the different performance characteristics of the storage technologies, they are best suited to supporting substantially different generation mixes. This indicates that the technical performance of energy storage has a significant impact on how the energy system should be sized.
Hu et al. \cite{hu2024OptimalPlanningElectricheating} considers a district energy system with local renewable generation providing both electricity and heat. It demonstrates that optimizing the sizing and location of battery storage, hydrogen storage, and CHP units in the system can reduce the net load fluctuation by 25\%. Similar benefits of using energy storage to manage variability in both renewable generation and load are shown by Phu et al. \cite{phu2024IGDTApproachMultiobjective}, which studies an energy park producing green hydrogen from biomass using electricity from wind and solar generation. Optimizing battery and hydrogen energy storage within the system is shown to substantially reduce the operating cost, carbon emissions, and grid reliance of the energy park. Further this optimized configuration is demonstrated to be robust against uncertainties in hydrogen demand, grid electricity price, and renewable generation.}

Many different energy storage technologies have been considered for \edit{supporting energy parks}, such as Lead-acid batteries, Lithium-ion batteries, compressed-air storage, and redox flow batteries \cite{irena2017ElectricityStorageRenewables}.
However the future performance of these technologies, i.e. how they will perform when implemented in the grid from around 2030 onwards, is still highly uncertain. Across the literature, estimates of technical parameters (e.g. round-trip efficiency and self-discharge rate) and economic parameters (e.g. cost per energy capacity and service lifetime) vary over wide ranges \edittwo{\cite{neso2022PotentialElectricityStorage}}. %\cite{irena2017ElectricityStorageRenewables,kebede2022ComprehensiveReviewStationary,kittner2020ChapterGridscaleEnergy,neso2022PotentialElectricityStorage,petkov2020PowertohydrogenSeasonalEnergy}. - for thesis
% comment that there is more uncertainty for newer technologies?
For example, a literature review, \cite{kebede2022ComprehensiveReviewStationary}, found estimates for Sodium-Sulphur high-temperature batteries ranging from 70--90\% for round-trip efficiency, and 0--20\% for self-discharge rate. \edittwo{\cite{petkov2020PowertohydrogenSeasonalEnergy} found values for the capital cost of Li-ion batteries between 150 and 600 \euro/kWh, and operational costs between 1 and 5\%. And an} IRENA report, \cite{irena2017ElectricityStorageRenewables}, estimated the annualized energy capacity cost of compressed-air storage to be between 2 and 355 cents/kWh/yr.

These highly uncertain technical and economic performance parameters have a significant impact on the cost of developing and operating grid-scale storage in energy parks. Crucially, it also causes uncertainty in which storage technology will provide the lowest cost pathway to supporting variable renewable generation. As research and development of storage technologies progresses, uncertainty in their performance will decrease, and the best technology will emerge. However, to meet 2030 and 2040 carbon emissions reductions targets, energy parks must be designed and begin development in the near future.

This raises the following question, ``When designing an energy park, can the best energy storage technology be identified with the current level of uncertainty in performance? Or should developers keep their options open to potentially make a better choice when uncertainty has reduced?''
No existing studies have considered the impact of reduction in storage performance uncertainty on the design of energy parks and the selection of storage technologies. In fact, very few consider storage uncertainties during energy park design.

\subsection{Storage technology performance uncertainty in energy park design}

% How has uncertainty in technology performance in design been handled? Has there been any work looking at uncertainty reduction?
% - lots of studies look at impact of uncertainty on energy park design (and even more on operation, maybe finds some refs)
% - lots of studies considering uncertainty in generation, load, and prices (e.g. elec \& carbon) \cite{sheibani2018EnergyStorageSystem} (review)
% - discuss a few;
% - \cite{kim2024MultiperiodMultitimescaleStochastic}
% - \cite{bakke2016InvestmentElectricEnergy}
% - \cite{gabrielli2019RobustOptimalDesign}
% - however, very few consider uncertainty in storage performance; some do cost \cite{kim2024MultiperiodMultitimescaleStochastic,bakke2016InvestmentElectricEnergy,chadly2023TechnoeconomicAssessmentEnergy,coppitters2021RobustDesignOptimization}, but only two \cite{chadly2023TechnoeconomicAssessmentEnergy,coppitters2021RobustDesignOptimization} consider uncertainty in technical parameters
% - no studies have considered uncertainty reduction and its impact on decision making in energy parks
% - uncertainty reduction has been considered in wider engineering context (cite my papers)

The impact of uncertainty on the operation of energy parks, and so the cost of arbitraging energy to match variable generation with demand, has been widely studied. % \cite{kim2017RobustOperationEnergy,zhang2023OptimalEconomicProgramming,ji2025OptimalSchedulingParklevel,li2021OptimalOperationModel}. % control/operational papers studying uncertainties - list for thesis
\edittwo{For instance, \cite{kim2017RobustOperationEnergy} shows that accounting for load forecast uncertainty in control can reduce operating costs by 10\%, and \cite{ji2025OptimalSchedulingParklevel} develops a method for managing the risk introduced by uncertainty in load and solar generation during operation.}
Many previous articles have investigated the optimal sizing of generation and storage assets to maximize total profit in the presence of various uncertainties.
For example, Kim et al. \cite{kim2024MultiperiodMultitimescaleStochastic} consider the sizing of wind generation, battery storage, and electrolyser capacity in a micro-grid with local demands for both electricity and hydrogen. A bi-level optimization method is used to minimize the average cost of meeting the local energy demands, accounting for uncertainty in wind generation and demand patterns on operating costs, and uncertainty in the per capacity costs of the assets.
Gabrielli et al. \cite{gabrielli2019RobustOptimalDesign} develop a method for sizing a multi-energy system with solar generation, battery storage, heat pumps, and an electrolyser, so that the design is robust to uncertainty in weather conditions and the electricity and heating demands which must be met. % mention total cost including carbon cost?
Bakke at al. \cite{bakke2016InvestmentElectricEnergy} study the optimal level and timing of investment in stand-alone battery storage for providing energy arbitrage and ancillary grid services, considering uncertainty in both electricity prices (which determine operational revenue) and evolution in the price per capacity of battery storage.

Uncertainty in renewable generation \edittwo{\cite{gabrielli2019RobustOptimalDesign}}, % \cite{kim2024MultiperiodMultitimescaleStochastic,gabrielli2019RobustOptimalDesign,yang2024PlanningMicrogridsIndustrial,barros2022MultiObjectiveOptimizationSolar,sheibani2018EnergyStorageSystem,coppitters2021RobustDesignOptimization,mohammadi2022EffectMultiUncertainties,rehman2022SizingBatteryEnergy,keyvandarian2023OptimalSizingReliabilityconstrained}
energy demands \edittwo{\cite{mohammadi2022EffectMultiUncertainties}}, % \cite{kim2024MultiperiodMultitimescaleStochastic,gabrielli2019RobustOptimalDesign,yang2024PlanningMicrogridsIndustrial,sheibani2018EnergyStorageSystem,coppitters2021RobustDesignOptimization,mohammadi2022EffectMultiUncertainties,rehman2022SizingBatteryEnergy,keyvandarian2023OptimalSizingReliabilityconstrained,maia2021CertaintyUncertaintyStochastic}
and the price of grid electricity \edittwo{\cite{rehman2022SizingBatteryEnergy}} % \cite{bakke2016InvestmentElectricEnergy,sheibani2018EnergyStorageSystem,coppitters2021RobustDesignOptimization,mohammadi2022EffectMultiUncertainties,rehman2022SizingBatteryEnergy}
and carbon \cite{yang2024PlanningMicrogridsIndustrial} are commonly considered during energy park design.
However, few studies account for uncertainties related to storage technologies. Some have included uncertainty in the cost of energy storage, \edittwo{e.g. \cite{kim2024MultiperiodMultitimescaleStochastic} \& \cite{bakke2016InvestmentElectricEnergy}.} %\cite{kim2024MultiperiodMultitimescaleStochastic,bakke2016InvestmentElectricEnergy,chadly2023TechnoeconomicAssessmentEnergy,coppitters2021RobustDesignOptimization}. - for thesis
\edit{However only two studies could be found that consider uncertainty in the technical performance of storage technologies, \cite{coppitters2021RobustDesignOptimization} which accounted for uncertain cycle life and self-discharge rate, and \cite{chadly2023TechnoeconomicAssessmentEnergy} which considered uncertain round-trip efficiency.}

No existing works have investigated the impact that uncertainty reduction has on the process of energy park design, as has recently been studied in the context of \edit{building operation \cite{langtry2024RationalisingDataCollection} and district energy system design \cite{langtry2024QuantifyingBenefitLoad}}.
Additionally, in the broader energy systems literature, risk aversion is known to be an important feature of decision making \edittwo{\cite{pickering2019PracticalOptimisationDistrict}.} %\cite{pickering2019PracticalOptimisationDistrict,mu2022CVaRbasedRiskAssessment,tostado-veliz2024RiskaverseElectrolyserSizing}
\edittwo{Its effect on the selection and sizing of energy infrastructure in parks has been investigated in studies such as \cite{mu2022CVaRbasedRiskAssessment} and \cite{tostado-veliz2024RiskaverseElectrolyserSizing} respectively.
However, the impact risk aversion} has on the benefit of uncertainty reduction is yet to be studied.

\subsection{Storage technology selection in energy park design}

% How has the different technology options for energy parks been studied? Has there been any work combining uncertainty reduction and optionality?

% - some work considering different storage technology options in energy parks; discuss a few
% - \cite{mazzoni2019EnergyStorageTechnologies} compares Li-ion, Pb-acid, VRFB batteries in DES with solar, diesel, cooling, and battery \& thermal storage, uses deterministic optimization
% - \cite{zhang2019BalancingWindpowerFluctuation} compares H2 to Li-ion storage for supporting grid-scale wind generation, using stochastic optimization for sizing, importantly considers uncertainty in storage price and efficiency (H2 only)
% - \cite{mohseni2022QuantifyingEffectsForecast}, compares Li-ion, Pb-acid, NiCd, NaS batteries in micro-grid with solar, wind, hydro-generation, and battery storage, using stochastic optimization for sizing (no uncertainties in storage considered)
% - \cite{chadly2023TechnoeconomicAssessmentEnergy} compares storage options for buildings; \cite{murrant2018AssessingEnergyStorage} does a broader and more holistic comparison of energy system options for a local grid
% - storage technology options have been considered in larger-scale studies (e.g. \cite{sepulveda2021DesignSpaceLongduration}, PyPSA, etc. \cite{find})
% - however all studies are single point decisions, none consider the impact of uncertainty reduction on technology selection decision making

While the majority of energy park design studies consider only a single storage option, a few do compare multiple storage technologies to identify the option best suited to the energy system.
%%\cite{zhang2019BalancingWindpowerFluctuationmazzoni2019EnergyStorageTechnologies,mohseni2022QuantifyingEffectsForecast} <- for thesis
Zhang et al. \cite{zhang2019BalancingWindpowerFluctuation} compares the use of hydrogen storage and Lithium-ion batteries to manage the variability of wind generation and reduce curtailment. \cite{mazzoni2019EnergyStorageTechnologies} \& \cite{mohseni2022QuantifyingEffectsForecast} compare different battery storage technologies for supporting distributed energy systems with local renewable generation. They find that Lead-acid and Sodium-Sulphur batteries provide the lowest total operating cost for their respective energy systems. However, none of these studies account for uncertainty in the characteristics of the storage technologies during the design optimizations which are then compared. A comparison of battery technologies which does account for performance uncertainties is performed in the context of building energy systems in \cite{chadly2023TechnoeconomicAssessmentEnergy}.
% mention large-scale studies?

The main limitation of all these existing studies is that they consider the energy park design process, including both technology selection and sizing, as a static, single-point decision, where the entire design is determined up-front. This does not reflect the design \edit{process} of practical energy systems, where several stages of engineering design are performed progressively with increasing detail, and greater information regarding available components (which for instance may be provided during contracting).

As uncertainty in the performance of the available storage technology options reduces, the best technology choice (that with, for example, the lowest average cost under the remaining uncertainty) may change. As a result, retaining optionality over \edit{which} storage technology \edit{is ultimately used} through the design process \edit{could be} valuable, as it may be possible to reduce the overall cost of the energy system by selecting a better performing technology at a later stage. However, keeping this optionality and gathering improved information about the performance of each storage technology is costly. This raises the question of whether the benefits of retaining storage optionality are worth the cost.

\subsection{Research objectives \& novel contributions}

% Usual structure
% Research gap is intersection of uncertainty reduction and technology options, looking at importance of retaining optionality through design process - question not asked before

In the existing literature, no studies have investigated the impact that reducing uncertainty in energy storage technology performance has on the design of energy parks or the selection of storage technologies. As there is currently large uncertainty in the performance of prominent storage technologies, it is important to understand whether developers are able to commit to a single technology up-front, or whether its is beneficial to hedge their bets and have the option to choose the best technology when better information about performance is available.
% understanding of this impact is needed to determine whether it is worthwhile investing resources to allow system designs to be adapted as those uncertainties reduce.

This study uses the Value of Information analysis (VoI) framework to numerically answer the following questions, ``How valuable is retaining optionality in storage technology choice for the design of an energy park as uncertainty reduces? Is this benefit to design worth the cost of obtaining improved estimates of storage performance and retaining the option to change storage technology?''. It investigates this in the context of sizing wind, solar, and storage capacities for an illustrative energy park system, modelled off a real-world green hydrogen plant proposal in the Port of Rotterdam.
The main objectives of this study are:
\begin{itemize}
    \itemsep0pt
    \item Quantify the benefit of \edit{updating} the system design (sizings) after uncertainty in storage performance has been reduced % by R\&D
    \item Determine whether the benefit to design of retaining optionality in storage technology selection by speculatively developing multiple technologies is worth its cost
    \item Investigate how the level of uncertainty reduction and risk aversion impact whether design adaptation and storage optionality are worthwhile
\end{itemize}

This work is the first to investigate the impact of reducing uncertainty in storage technology performance on energy system design, and quantify the benefit of retaining optionality in storage \edit{technology} selection as uncertainty reduces. It demonstrates the value to energy system developers of providing the opportunity to \edit{update} system designs as improved information about the performance of energy storage (with reduced uncertainty) becomes available, which has important implications for industry design practice.

The remainder of this work is structured as follows.
Section \ref{sec:method} outlines the Value of Information analysis framework used to study design under uncertainty reduction.
Section \ref{sec:experiment} describes the illustrative energy park studied, including the probabilistic models of renewable generation, industrial load, and energy storage technology performance uncertainty used, and the Stochastic Programming model used to perform system design.
Section \ref{sec:results} presents the results of the numerical experiments and discusses their importance for informing energy park design practices. An initial experiment quantifies the value of being able to update the energy park design, and change choice of energy storage technology, after uncertainty in storage technology performance has been reduced by R\&D. However, as no information is available in the literature regarding how much R\&D might reduce uncertainty, i.e. how closely the performance of a demonstrator system matches a large-scale version, a sensitivity analysis over the level of uncertainty reduction is performed. The Value of Optionality is then investigated in two further cases where two energy storage technologies are used in the park, and where risk aversion is incorporated into the design process.
Finally conclusions are drawn in Section \ref{sec:conclusions}.
\newpage
\section{Methodology} \label{sec:method}

% Aim for sub 1000 words
% Focus on Bayesian decision analysis as relevant theory; mention VoI with less focus, and use it to disucss VoO as related idea

\edit{Value of Information analysis, originally proposed by Raiffa \cite{raiffa1969ReviewDecisionAnalysis} and Howard \cite{howard1966InformationValueTheory} in the 1960s, is a framework based on Bayesian Decision Analysis and Expected Utility Theory \cite{smith1945TheoryGamesEconomic} for quantifying the improvement in decision making provided by uncertainty reduction.
This section briefly outlines the methodology, and shows how it can be extended to study the benefit of decision optionality as uncertainty reduces.}

\subsection{Bayesian Decision Analysis} \label{sec:bayes-theory}

Bayesian decision analysis provides a mathematical framework for studying decision-making in the presence of uncertainties, referred to as stochastic decision problems. Its aim is to determine the optimal set of actions which should be taken by a decision-maker (termed an `actor') in order to maximise their expected utility. That is, find the decision which when taken in the system provides the highest reward/benefit to the decision-maker on average over the uncertainties in the problem. This task can be formulated as a mathematical (stochastic) optimization problem.

Consider a generalised stochastic decision problem in which an actor seeks to select a `decision action' to take, $a \in \mathcal{A}$, within a system with uncertain parameters $\theta$, which have a prior probabilistic model (distribution), $\pi(\theta)$. The performance/benefit of each available action is described by a utility function which is also dependent upon the uncertain parameters, $u(a,\theta)$. In Bayesian decision analysis, before an action $a$ is taken, the actor may choose to take a `measurement action', $e \in E$, which provides some data $z$ that reduces the uncertainty in $\theta$. The probabilistic model describing the measurement data $f_e(z|\theta)$ is used to update the prior model, $\pi(\theta)$, to produce a posterior probabilistic model (distribution), $\pi(\theta|z)$. This posterior (which has reduced uncertainty c.f. the prior) is then used by the actor to inform their choice of `decision action', improving their decision making performance.

\begin{figure}[h]
    \centering
    \includegraphics[width=0.8\linewidth]{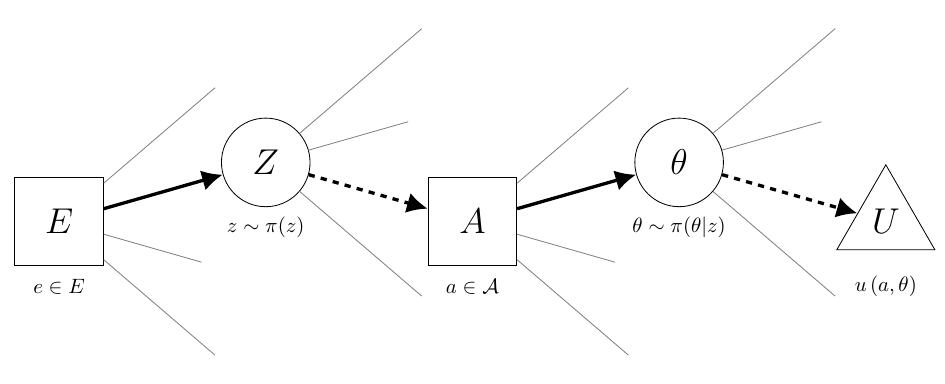}
    \vspace*{-0.5cm}
    \caption{Decision tree representation of Pre-Posterior Decision Problem}
    \label{fig:DT-prepost}
\end{figure}

The set of available actions, prior probabilistic model, and utility function, $\lbrace\mathcal{A},\pi(\theta),u(a,\theta)\rbrace$, provide a complete mathematical description of the decision making task under uncertainty. The likelihood function $f_e(z|\theta)$ describes the reduction in epistemic uncertainty in the parameters of the system, $\theta$, provided by data collection.
This generalised model can be represented graphically in decision tree form, as shown in Fig. \ref{fig:DT-prepost}, in which square nodes represent decisions, circular nodes represent uncertainties, and triangular nodes represent utilities.

The actor, who is assumed to be risk neutral, aims to maximise the expected utility they receive from the `decision action' $a$ they select. Costs are defined as negative utilities. The actor may choose to do this without taking any measurement. The resulting stochastic optimisation is termed the Prior Decision Problem,
\begin{equation} \label{eq:u-prior}
    \max_{a \in \mathcal{A}} \, \mathbb{E}_{\theta} \left\lbrace u(a,\theta) \right\rbrace
\end{equation}

Alternatively, the actor can initially take a `measurement action' $e$ which reduces uncertainty in the problem, and improves their subsequent choice of `decision action'. In this case, the optimization of expected utility is performed over both measurement and decision actions, and is termed the Pre-Posterior Decision Problem,
\begin{equation} \label{eq:u-prepost}
    \max_{e \in E} \, \mathbb{E}_{z} \left\lbrace \max_{a \in \mathcal{A}} \, \mathbb{E}_{\theta|z} \left\lbrace u(a,\theta) \right\rbrace \right\rbrace
\end{equation}
\vspace{0.1cm}

\subsection{Value of Information} \label{sec:voi}

By taking a measurement and reducing uncertainty in the problem the actor is able to improve their decision making, i.e. increase the average utility they obtain from the decision problem. This increase in expected utility provided by uncertainty reduction from measurement (the difference in expected utilities achieved when decisions are made with and without uncertainty reduction respectively) is termed the Value of Information (VoI) \citep{raiffa1969ReviewDecisionAnalysis}.

If a measurement $e$ is taken which provides imperfect/uncertain information that reduces but does not remove the epistemic uncertainty in the parameters $\theta$, the expected value of that uncertain information, termed the Expected Value of Imperfect Information (EVII), is computed as,
\begin{equation} \label{eq:EVII}
    \text{EVII}(e) = \mathbb{E}_{z} \left\lbrace \max_{a \in \mathcal{A}} \: \mathbb{E}_{\theta|z} \left\lbrace u(a,\theta) \right\rbrace \right\rbrace - \max_{a \in \mathcal{A}} \: \mathbb{E}_{\theta} \left\lbrace u(a,\theta) \right\rbrace
\end{equation}
where the posterior probabilistic model, $\pi(\theta|z)$, used to compute inner expectation with respect to $\theta|z$, is derived using the likelihood function for the measurement $e$, $f_e(z|\theta)$.

% In the case where measurement provides perfect information on the true value of the uncertain parameters of the system, the expected utility obtained by the actor making decisions with perfect information simplifies as,
% \begin{equation}
% \mathbb{E}_{z} \left\lbrace \max_{a \in \mathcal{A}} \, \mathbb{E}_{\theta|z} \left\lbrace u(a,\theta) \right\rbrace \right\rbrace \: \longrightarrow \: \mathbb{E}_{z} \left\lbrace \max_{a \in \mathcal{A}} \, u(a,z) \right\rbrace
% \end{equation}
% The expected value of information in this case is termed the Expected Value of Perfect Information (EVPI), and is given by,
% \begin{equation} \label{eq:EVPI}
%     \begin{aligned}
%         \text{EVPI} &= \mathbb{E}_{z} \left\lbrace \max_{a \in \mathcal{A}} \, u(a,z) \right\rbrace - \max_{a \in \mathcal{A}} \mathbb{E}_{\theta} \left\lbrace u(a,\theta) \right\rbrace \\[1em]
%     \end{aligned}
% \end{equation}
% The EVPI is significantly cheaper to compute, and provides an upper-bound on the EVII \cite{keisler2014ValueInformationAnalysis}.\\

The Value of Information quantifies how much better on average the decision-maker is able to do at making their decision by reducing uncertainty. Typically, decision problems are formulated using an economic objective, i.e. a total cost or profit, and so the VoI quantifies the actor's willingness to pay to reduce uncertainty at the time the decision must be taken. Comparing the VoI to the cost of reducing uncertainty allows its net economic benefit to be quantified, i.e. the VoI minus the cost of information. With this, a decision maker can determine whether a measurement/uncertainty reduction is economically worthwhile, and compare the relative benefit of different uncertainty reduction options.
% NOTE: for the VoO the utility is nicely defined (wrapping up SP), and the tech selection is an exact choice so the theory is simple, but the VoI being quantified is actually On-Policy VoI (make comment and cite)

\subsection{Value of Optionality} \label{sec:voo}

% SDP defined by {A,pi,u} - u determines the costs so can't be changed (changes problem), but the two remaining things can change to improve outcomes, pi and A (VoI and VoO resp.)

A stochastic decision problem is defined by: a set of available actions, $\mathcal{A}$, a probabilistic model of the uncertain parameters in the system, $\pi(\theta)$, and a utility function describing the behaviour of the system and the performance/benefit that results when an action is taken in the system, $u(a,\theta)$. The utility function defines the problem being studied and so cannot be changed. However, analogously to the Value of Information (VoI), the benefit to the decision-maker of improving the choice of actions available to them can be studied and quantified.
This is termed the Value of Optionality (VoO).

Consider the case where the set of `decision actions' available to the actor is expanded, to $\widehat{\mathcal{A}} \supset \mathcal{A}$, at the same time as uncertainty is reduced by taking a measurement, $e$. The Expected Value of Optionality (EVO) is found by comparing the expected utility achieved in the case with increased optionality to the original case, 

\begin{equation} \label{eq:EVO}
    \begin{aligned}
        \text{EVO}(\widehat{\mathcal{A}},e) &= \mathbb{E}_{z} \bigg\lbrace \max_{a \in \widehat{\mathcal{A}}} \: \mathbb{E}_{\theta|z} \left\lbrace u(a,\theta) \right\rbrace \bigg\rbrace - \mathbb{E}_{z} \bigg\lbrace \max_{a \in \mathcal{A}} \: \mathbb{E}_{\theta|z} \left\lbrace u(a,\theta) \right\rbrace \bigg\rbrace \\[1ex]
        &= \mathbb{E}_{z} \left\lbrace \max_{a \in \widehat{\mathcal{A}}} \: \mathbb{E}_{\theta|z} \left\lbrace u(a,\theta) \right\rbrace - \max_{a \in \mathcal{A}} \: \mathbb{E}_{\theta|z} \left\lbrace u(a,\theta) \right\rbrace \right\rbrace \\[1ex]
        % &= \mathbb{E}_{z} \left\lbrace \:
        %     \begin{aligned}
        %     & \mathbb{E}_{\theta|z} \left\lbrace u(a^*_z,\theta) \right\rbrace - ... \ & \text{if} \ \ a^*_z \notin \mathcal{A} \\
        %     & \qquad\quad 0 \ & \text{otherwise} \\
        %     \end{aligned}
        % \: \right\rbrace
    \end{aligned}
\end{equation}
noting that the inner term is zero unless the optimal action for the posterior (when data $z$ is measured) is not in original action set.\\

Similarly to the VoI, the Value of Optionality quantifies the average improvement in decision-making provided by additional optionality, and so the actor's willingness to pay for having a greater choice of actions at the time the decision must be taken. However, there is often a cost associated with having actions available, even if they are not taken. Comparing the VoO to the cost of increasing the decision-maker's options determines whether increasing optionality is worthwhile.

% an accessible example of this is determining the number of sizes of box a distribution centre (e.g. Amazon) should stock in its packing system, given it is both costly to have a greater stock of boxes, and costly to pad out a large box to fit a smaller item
% note, there needs to be a cost associated with having options available but not necessarily used
\newpage
\section{Experimental setup} \label{sec:experiment}

% Aim for 1500 words

%\subsection{Illustrative energy park system} \label{sec:setup}

% Explain energy system, motivation (interest in energy parks \& hydrogen context), location, data used, etc.
% What energy system are we studying and why?

The value of retaining optionality in energy storage technology choice during design is investigated for an illustrative energy park system. This system is modelled on a real-world energy park proposal. Due to its importance for decarbonizing transportation, heavy industry, and chemical manufacturing, there has been substantial commercial interest in the production of low-carbon hydrogen at scale. A proposal to build a \SI{250}{MW} hydrogen electrolyser facility in the Port of Rotterdam to decarbonize industry and transportation in the area has received early-stage support from the Dutch government \cite{bp2022GreenHydrogenProject}. % \cite{2022ProjectH2Fifty,bp2022GreenHydrogenProject}
Co-locating renewable generation and energy storage with this industrial hydrogen plant in an energy park would reduce the total cost of operating the system, and its impact on the electricity grid.

\subsection{System design under uncertainty}

\begin{figure}[!b]
    \centering
    \includegraphics[width=0.65\linewidth]{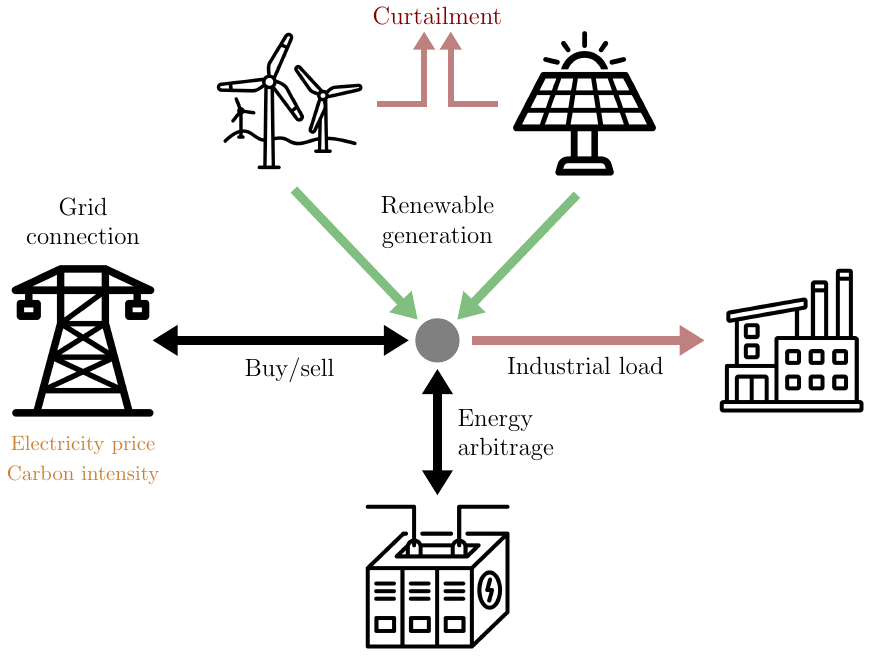}
    \caption{Schematic of energy flows within energy park model}
    \label{fig:system-diagram}
\end{figure}

A key design challenge is determining the sizing of generation and storage, and the energy storage technology, that maximize the profitability of the energy park. It is assumed that offshore wind and solar generation (via direct wire connection) can be installed to support the site, and that energy can be traded with the grid. Fig. \ref{fig:system-diagram} illustrates the energy flows within the system model. Limits are imposed on grid connection capacity to reflect network restrictions, and on the maximum solar generation capacity to reflect land scarcity. Additionally, a capital budget constraint is imposed. Optimization-based design is used to select the energy storage technology and generation \& storage sizings that minimize the expected cost of providing the industrial load. A carbon price is included to reflect the incentives for low-carbon hydrogen production. Uncertainties in the industrial load, patterns of wind and solar generation, and cost and technical performance of the different storage technologies are accounted for during design.

% Explain decision sequence, what it corresponds to, why it's important, what we're studying and why (maybe create a sequence diagram to illustrate decision making process; initial design, R\&D, final design, operation); explain R\&D option and associated cost to project
% What is the impact of uncertainty on the design problem and why is thinking about optionality important? How does it fit into the problem?

\begin{figure}[h]
    \centering
    \includegraphics[width=\linewidth]{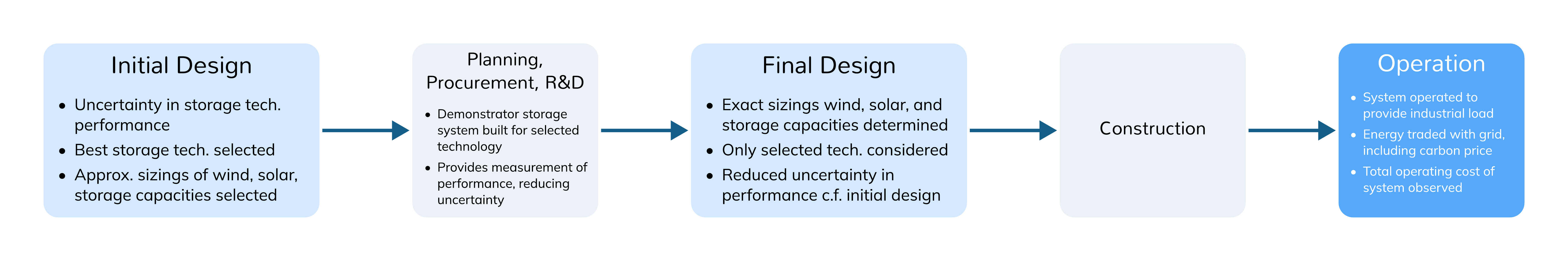}
    \caption{Two-stage decision model of energy park design}
    \label{fig:decision-sequence}
\end{figure}

A two-stage decision model of the energy park design process is used. In an initial design stage, an energy storage technology is selected, and approximate sizings of the wind, solar, and storage capacities are determined (for consenting, planning permission, and procurement).
At the time of initial design, the expected performance of each storage technology during operation is uncertain. As project planning and the pre-construction phases are completed, it is assumed that procurement and R\&D is undertaken for the chosen storage technology, leading to the creation of a small-scale demonstrator. This demonstrator storage system provides the energy park designer with measurements of the storage technology performance, which reduce uncertainty in the actual performance of a large-scale system. However, performing this procurement and R\&D to reduce storage uncertainty is costly.
After the demonstrator project is complete, a final design stage selects the exact wind, solar, and storage capacities for the energy park system. Finally, the system is constructed and operated, and the overall cost of providing the industrial load is observed.
This decision model is illustrated in Fig. \ref{fig:decision-sequence}.

% raises question; should R\&D be performed for multiple technologies even if only one is eventually used? is the optionality in the 2nd stage worthwhile?

This decision model raises a question, ``Even though only one storage technology is ultimately used in the energy park, is it worthwhile performing procurement and R\&D for all \edit{available} technologies, so that at the final design stage, a possibly better choice of storage technology can be made once uncertainty has been reduced?'', i.e. ``Is retaining optionality in storage technology choice until the final design stage worth the cost?''.

\subsection{\edit{Probabilistic models of renewable generation \& industrial load}} \label{sec:prob-gen-load}

% Use tables for dist. params for each storage technology, and prob. model for other uncertainties
% Include plots of time series uncertainties, e.g. wind \& solar generation

%\subsubsection{Renewable generation \& industrial load}

% Wind generation \& solar generation (historic data, and specify locations assumed), load level ($\pm20\%$ with curtailment to prevent extreme values)

\begin{figure}[h]
    \centering
    \subfloat[Summer.]{
        \includegraphics[width=\linewidth]{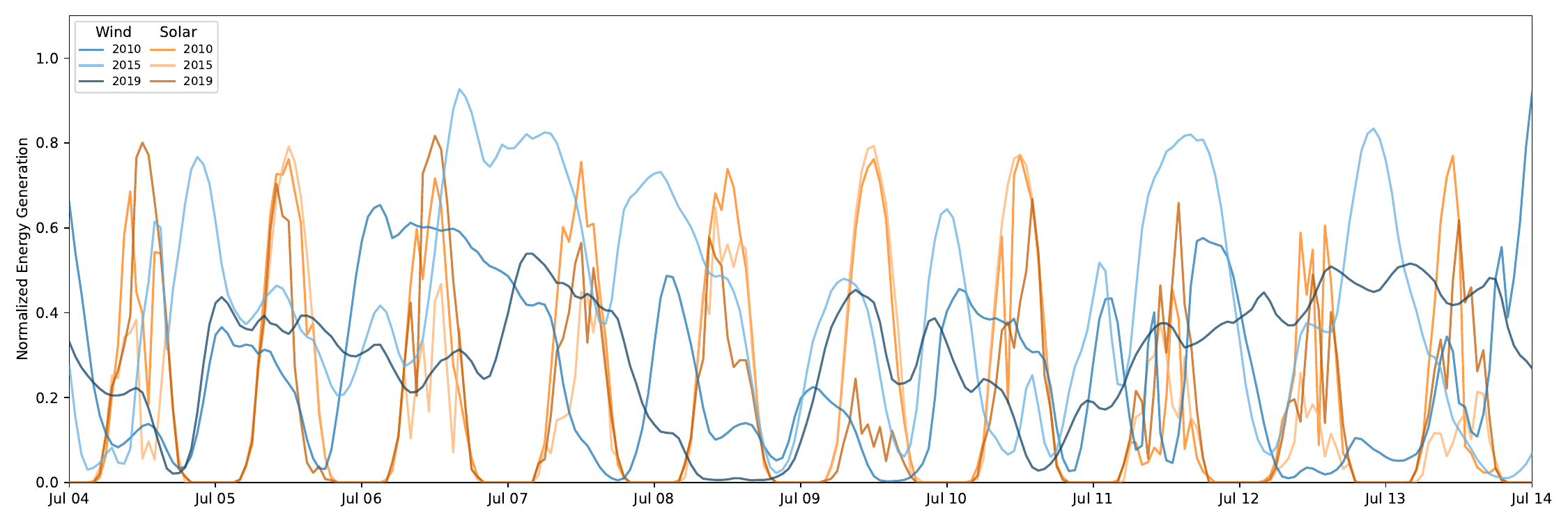} \label{fig:summer-gen}
    }
    \vspace{-0.1cm}
    \subfloat[Winter.]{
        \includegraphics[width=\linewidth]{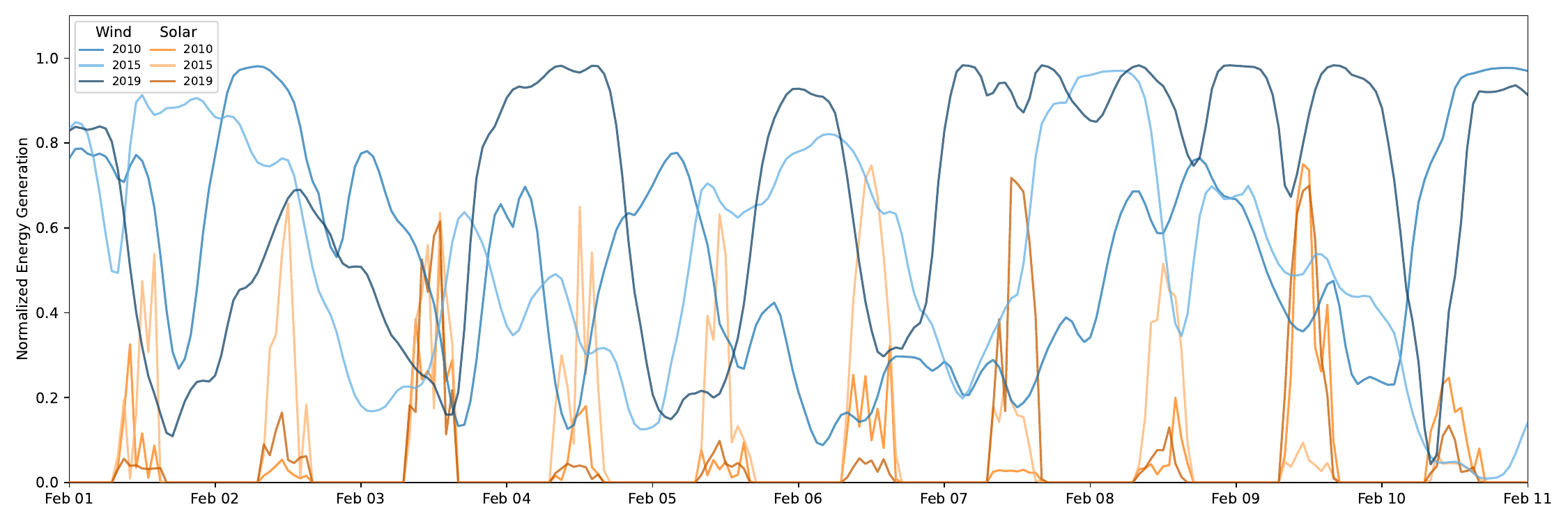} \label{fig:winter-gen}
    }
    \vspace*{-0.2cm}
    \caption{Examples of wind and solar normalized generation profiles for Rotterdam.} \label{fig:example-gen-data}
    \vspace{0.1cm}
    \small{\textit{The complete dataset of wind and solar generation profiles used in the case study can be view interactively} \href{https://mal84emma.github.io/Energy-Park-Design/generation_dataset_plot.html}{\textbf{here}}.}
\end{figure}

Historic data is used to construct probabilistic models of uncertain wind and solar generation. In both cases, generation potential per installed capacity data was \edit{gathered} for 2010-2019. Uncertainty in renewable generation is modelled by randomly sampling a year from that range and using the corresponding generation data. For wind, data was collected from the \href{https://www.renewables.ninja/}{renewables.ninja} model \edittwo{\cite{staffell2016Renewablesninja}} at the location of a wind farm development zone near Rotterdam (IJmuiden Ver), using a power model from a typical offshore wind turbine. For solar, data was collected from the EU's \href{https://joint-research-centre.ec.europa.eu/photovoltaic-geographical-information-system-pvgis_en}{PVGIS} model \cite{jrceuproeancommission2017JRCPhotovoltaicGeographical}, using default PV module settings.
Fig. \ref{fig:example-gen-data} plots this data from periods in both summer and winter for three example years.

\newpage

Uncertainty in the demand for hydrogen in the local market during operation \edit{leads to} uncertainty in the \edit{final sizing of the electrolyser. The conversion efficiency of the electrolyser units also has some uncertainty. So, at the time the supporting generation and storage infrastructure is designed, the electrical load that will be required by the hydrogen plant is uncertain.}
\edit{The industrial load} is modelled as being constant during operation\edit{, as electrolysers are typically run at high capacity factors to provide a competitive hydrogen price \cite{furfari2021GreenHydrogenCrucial}, but} with an unknown\edit{/uncertain} level. A truncated Gaussian distribution \edit{ is used for the load, with} mean set to the proposed electrolyser capacity of \SI{250}{MW}, \edit{and} standard deviation $\sigma$ taken to be \SI{25}{MW} (10\%), \edit{truncated} at $\pm2\sigma$. % to limit unrealistic extreme values

The probabilistic models of the renewable generation and industrial load time series are summarized in Table \ref{tab:ts-prob-models}.
\edit{Grid electricity price and carbon intensity data for the Netherlands from 2023 is used as a base case}\footnotemark. % most representative data available
Uncertainties in price and carbon intensity \edit{are} not modelled\edit{, as due to global energy market events, few years of representative data are available. However 2024 data is used to perform a sensitivity analysis to determine if changes in grid price and carbon impact the conclusions derived.}\\
\footnotetext{Further information on data collection and processing is available \href{https://github.com/mal84emma/Energy-Park-Design/blob/main/data/data_sources.md}{online}.}

\begin{table}[h]
    \centering
    \renewcommand{\arraystretch}{1.5}
    \setlength{\tabcolsep}{12pt}
    \begin{tabular}{c|ccc} \toprule \toprule
        Parameter & Distribution & Parameters & Info \\
        \midrule \midrule
        Wind & \multirow{2}{*}{Discrete uniform} & \multirow{2}{*}{$\{ 2010,\ldots,2019 \}$} & \multirow{2}{*}{\small{\shortstack[l]{Years of historic generation\\potential data sampled}}} \\
        Solar & &  & \\
        Load & Truncated Gaussian & \makecell{$\mu=\SI{250}{MW}$\\$\sigma=\SI{25}{MW}$\\$\text{cut-off} = \pm2\sigma$} & \small{\shortstack[l]{Truncation used to limit\\unrealistic extreme load values}} \\
        \bottomrule \bottomrule
    \end{tabular}
    \caption{Probabilistic models of renewable generation and industrial load}
    \label{tab:ts-prob-models}
\end{table}
\vspace*{-0.4cm}

%\subsubsection{Storage technology performance \& measurement}
\subsection{\edit{Probabilistic models of storage technology performance \& measurement}} \label{sec:prob-storage}

% Detail probabilistic models of stuff used, where they're derived from, and what they correspond to (interpretation of prior and measurement models, including $r$ param; understanding of technology performance \& demonstrator project, $r$ models confidence in demonstrator)

At the time the energy park is designed, the actual performance of the available storage technologies during operation (with regards cost and technical aspects such as efficiency) is unknown/uncertain. However, the energy system designer can develop a probabilistic understanding of how each storage technology might perform from existing storage systems (if any exist) and estimates from the scientific literature.

This study considers four candidate energy storage technologies for use in the energy park:
\begin{itemize}
\itemsep0ex
    \item Lithium ion batteries (Li-ion)
    \item Sodium-Sulphur high-temperature batteries (NaS)
    \item Vanadium redox-flow batteries (VRFB)
    \item Compressed-air energy storage (CAES)
\end{itemize}
and considers uncertainty in the following economic and technical performance parameters of each storage technology:
\begin{itemize}
\itemsep0ex
    \item Cost per unit energy capacity (\euro/kWh)
    \item System lifetime (years)
    \item Round-trip efficiency (\%)
\end{itemize}
It is assumed that the depth-of-discharge and discharge ratio\footnotemark of each technology are known, with values taken from \cite{irena2017ElectricityStorageRenewables} and \cite{kebede2022ComprehensiveReviewStationary} respectively.
% not enough information in literature to get a decent prior distribution, and keeps the analysis a bit cleaner
\footnotetext{The ratio of power capacity to energy capacity, also frequently referred to using the `storage duration' (how long the storage unit can discharge for at full power output) which is the reciprocal.}

The level of uncertainty varies both between different technologies, due to varying levels of technology maturity and research, and between performance parameters, due to the underlying factors causing uncertainty, such as raw material prices. % for simplicity same shape of distribution used for all

The uncertainty in each parameter for each technology at the time of initial design before any R\&D has been performed, is modelled using a Gaussian distribution truncated at two standard deviations. This is referred to as the prior probabilistic model,

\begin{equation} \label{eq:prior}
    \theta \sim TN\left(\mu,\sigma,2\sigma\right)
\end{equation}
All uncertain parameters are taken to be independent, with mean ($\mu$) and standard deviation ($\sigma$) values derived separately for each parameter from literature estimates. % \cite{irena2017ElectricityStorageRenewables,kebede2022ComprehensiveReviewStationary,petkov2020PowertohydrogenSeasonalEnergy,kittner2020ChapterGridscaleEnergy}. % other references were used for cross-checking/validation - for thesis

When R\&D is performed and the demonstrator system is built and tested, this provides the energy park designer with a measurement of the performance of that storage technology. This measurement can be used to update their understanding of how a large-scale storage system for that technology will perform. The information provided by this measurement is combined with the prior probabilistic model to produce a posterior distribution describing the remaining (reduced) uncertainty.
The probability of obtaining a measurement, $z$, from the demonstrator system given the true performance of a large-scale storage system, $\theta$, called the likelihood model, is taken to be,
\begin{equation} \label{eq:measurement}
    z|\theta \sim \mathcal{N}\left(\theta,r\sigma\right)
\end{equation}
The uncertainty reduction factor, $r$, models the confidence that the energy park designer has in the demonstrator, i.e. how closely they believe the performance of the demonstrator reflects that of a large-scale system. \edit{Determining a precise numerical value for this parameter is challenging, as there is very little data available regarding how the performance of demonstrator systems compares to large-scale energy storage plants, particularly in the case of newer technologies such as CAES and VRFB where few large-scale projects have been completed. An initial estimate of $r=0.25$ is used for the experiments, representing a case where the designer has good confidence in the demonstrator, but retains a reasonable fraction of the initial uncertainty. To overcome this limitation, a sensitivity analysis over the value of $r$ is performed. Note that because $r$ is a multiplicative factor, the more mature storage technologies with lower uncertainty in their prior distributions also have correspondingly lower uncertainties in the posteriors.}\\

\begin{figure}[!ht]
    \centering
    \includegraphics[width=0.65\linewidth]{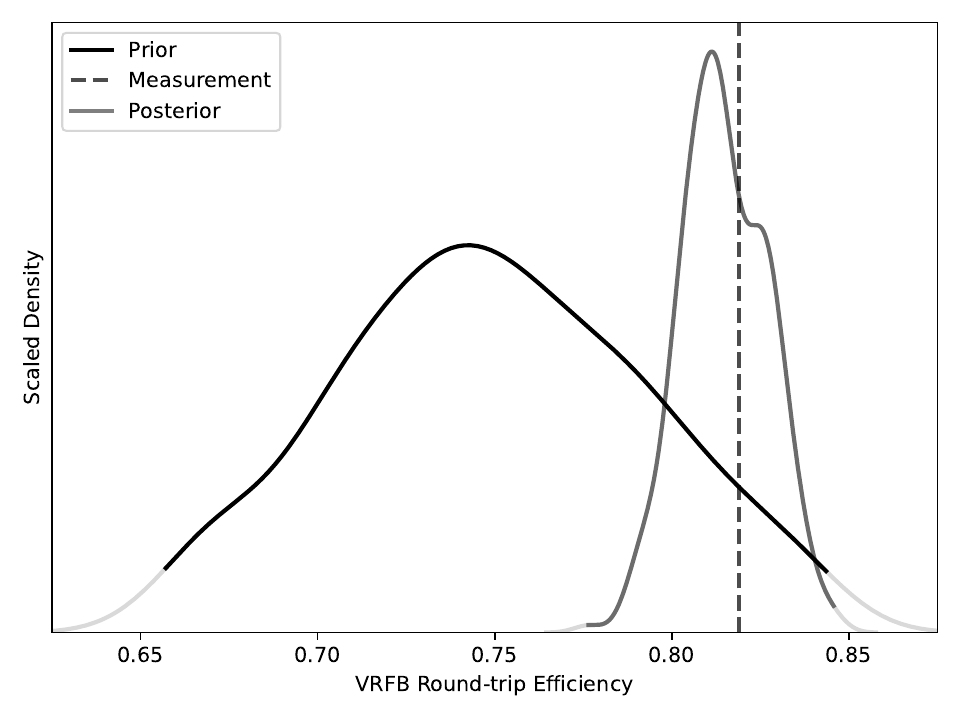}
    \caption{Prior distribution of storage efficiency, and corresponding posterior for an example measurement value.}
    \label{fig:msr-dists}
    \vspace{0.1cm}
    \footnotesize{\textit{Note: distributions (kernel density estimates) are not quite truncated Gaussian due to small number of samples used (250); densities have been scaled to allow the distributions to be visually compared.}}
\end{figure}

Table \ref{tab:tech-prob-models} summarises the probabilistic models of the uncertain performance of large-scale storage systems, and the measurements obtained from the demonstrators, providing mean and standard deviation values for each uncertain parameter.
Fig. \ref{fig:msr-dists} illustrates the prior distribution for the round-trip efficiency of VRFB storage, and the corresponding posterior distribution for an example measurement from a demonstrator system, showing the reduction in uncertainty provided by the demonstrator.\\

\begin{table}[!ht]
    \centering
    \renewcommand{\arraystretch}{1}
    \begin{tabularx}{0.9\linewidth}{YYYYYY} \toprule \toprule
        \multicolumn{6}{c}{\small Probabilistic models} \\[1ex]
        Prior & \multicolumn{4}{c}{$\theta \sim TN\left(\mu,\sigma,2\sigma\right)$} & \\[1ex]
        Likelihood & \multicolumn{4}{c}{$z|\theta \sim \mathcal{N}\left(\theta,r\sigma\right)$} & r=0.25 \\[1ex]
        & \multicolumn{4}{c}{\small\it Independent for each parameter \& technology} & \\
        \midrule \midrule
        \multicolumn{6}{c}{\small Model parameters} \\
         Parameter & Units & Technology & Mean $\mu$ & Std. dev. $\sigma$ & References \\
        \midrule
        \multirow{4}{*}{Cost} & \multirow{4}{*}{\euro/kWh} & Li-ion & 200 & 50 & \multirow{4}{*}{\cite{irena2017ElectricityStorageRenewables}} \\
        & & NaS & 175 & 37.5 & \\
        & & VRFB & 250 & 75 & \\
        & & CAES & 50 & 15 & \\
        \arrayrulecolor{black!50}\midrule
        \multirow{4}{*}{Lifetime} & \multirow{4}{*}{years} & Li-ion & 20 & 5 & \multirow{4}{*}{\cite{kebede2022ComprehensiveReviewStationary}} \\
        & & NaS & 25 & 5 & \\
        & & VRFB & 20 & 5 & \\
        & & CAES & 25 & 2.5 & \\
        \arrayrulecolor{black!50}\midrule
        \multirow{4}{*}{Efficiency} & \multirow{4}{*}{\%} & Li-ion & 92 & 3.5 & \multirow{4}{*}{\cite{kebede2022ComprehensiveReviewStationary}} \\
        & & NaS & 80 & 5 & \\
        & & VRFB & 75 & 5 & \\
        & & CAES & 60 & 2.5 & \\
        \bottomrule \bottomrule
    \end{tabularx}
    \caption{Probabilistic models of uncertain storage technology performance parameters and their measurement.} \label{tab:tech-prob-models}
\end{table}

\newpage

\subsection{Stochastic Programming for system design} \label{sec:SP}

% Detail SP model used to make decisions (standard in lit., include refs)
%? refer to On-Policy VoI paper - note that as decision is made using SP, we actually quantify On-Policy VoI
% include scenario reduction scheme \cite{heitsch2003ScenarioReductionAlgorithms,gioia2023ScenarioReducer}
% provide settings in appendix as in BD-VOI

Stochastic Programming is used to optimize the capacities of wind and solar generation and energy storage in the energy park system. Specifically, linear scenario programming is used due to its computational efficiency, enabling the use of an hourly resolved model with multiple scenarios, and its resulting prevalent use in the literature for designing district- and national-scale energy systems, see  \cite{pickering2019DistrictEnergySystem} \& \cite{yue2018ReviewApproachesUncertainty}.
%\cite{decarolis2017FormalizingBestPractice,pickering2019DistrictEnergySystem,yue2018ReviewApproachesUncertainty,2024OpenEnergyModelling} - for thesis

A linearized model of energy flows within the park system (refer to Fig. \ref{fig:system-diagram}) is used. The system design optimization task is formulated as a Linear Program\footnotemark. Eq. \ref{eq:SP-formualtion} describes this formulation for general case where multiple storage technologies are used together in the energy park. Table \ref{tab:SP-params} provides descriptions of all model parameters. To determine the best storage technology (or combination), the optimization is solved for each option independently, and the lowest cost setup is identified.
\footnotetext{As the design optimization is solved approximately using a Linear Program, estimates of the VoI and VoO are provided. The accuracy of such estimates is discussed in \cite{langtry2024QuantifyingBenefitLoad}.}

The design optimization aims to minimize the expected annualized cost of the energy park, the average total cost of running the energy park to provide the industrial load over all scenarios considered. The total cost is made up of the capital costs of the wind, solar, and energy storage, the cost (or profit) of energy bought from/sold to the grid, and the cost of carbon emissions associated with purchased grid electricity.
The optimization is subject to energy conservation\footnote{$[\,\cdot\,]^+$ and $[\,\cdot\,]^-$ represent the positive and negative parts of the argument respectively.} (Eq. \ref{eq:dynamics-constraint}), storage capacity (Eq. \ref{eq:power-constraint} \& Eq. \ref{eq:energy-constraint}), and grid capacity (Eq. \ref{eq:grid-cap}) constraints for operation within each scenario $m$. As well as a budget constraint on total capital expenditure (Eq. \ref{eq:budget-cap}). During operation, electricity can be bought from and sold to the electricity grid (Eq. \ref{eq:grid-variable}), and the renewable generation can be dynamically curtailed (Eq. \ref{eq:curtailment}).\\

Scenarios are generated by sampling from the distributions of renewable generation, industrial load, and storage technology performance described in Sec. \ref{sec:prob-gen-load} \& \ref{sec:prob-storage}. The sampled storage capacity cost and lifetime values are combined to produce annualized capacity cost samples for each technology, $p^s_{i,m}$.
The number of scenarios that can be considered during optimization is limited by high computational cost. Therefore, scenario reduction \cite{pickering2019PracticalOptimisationDistrict} is used\footnote{\edittwo{Specifically the fast-forward scenario reduction algorithm from \cite{heitsch2003ScenarioReductionAlgorithms}, implemented by \cite{gioia2023ScenarioReducer}.}} to improve the statistical representation of possible scenarios in the optimization. A large initial sample of scenarios is drawn, the optimization is performed considering each scenario separately, and a subset of scenarios which best represent the distribution of individually optimized costs is selected for use in the Stochastic Program.

The parameter settings used across all experiments in this study are detailed in \ref{app:case-study-params}.

\newcommand{\eqnskip}{1ex}
\newcommand{\smalleqnskip}{-1ex}
\renewcommand{\arraystretch}{1}
\begin{subequations} \label{eq:SP-formualtion}
    \begin{align}
        \addtocounter{equation}{-1}
        \mathlarger{\min} & \qquad p^{\textrm{w}} C^{\textrm{w}} + p^{\textrm{pv}} C^{\textrm{pv}} + \mathlarger{\mathlarger{\sum}}_m \: \rho_m \left( \mathlarger{\sum}_i \: p^s_{i,m} C^s_i + \mathlarger{\sum}_t \left( p^e[t] E^{\text{grid}}_m[t] + \, p^c \, c[t] \left[E^{\text{grid}}_m[t]\right]^+ \right) \right)
        \label{eq:lp} \\[\eqnskip]%%
        \text{over} & \qquad C^{\textrm{w}},\, C^{\textrm{pv}},\, C^s_i,\, \nabla_m[t],\,  E_{i,m}[t],\, \textrm{SoC}_{i,m}[t{+}1] \quad \forall \: i,\, m,\, t \tag*{} \\[\eqnskip]%% z_i
        \text{subject to} & \qquad \textrm{SoC}_{i,m}[t{+}1] = \textrm{SoC}_{i,m}[t] + \sqrt{\eta_{i,m}} \left[E_{i,m}[t]\right]^{+} - 1/\sqrt{\eta_{i,m}} \left[\, E_{i,m}[t] \right]^{-} \label{eq:dynamics-constraint} \\[\smalleqnskip]
        & \qquad -P^{\textrm{max}}_i \Delta t \leq E_{i,m}[t] \leq P^{\textrm{max}}_i \Delta t \label{eq:power-constraint} \\[\smalleqnskip]
        & \qquad (1{-}\nu_i)\, C^s_i \leq \textrm{SoC}_{i,m}[t{+}1] \leq C^s_i \label{eq:energy-constraint} \\[\smalleqnskip]
        & \qquad P^{\textrm{max}}_i = \delta_i C^s_i \label{eq:power-capacity} \\[\smalleqnskip]
        & \qquad \textrm{SoC}_{i,m}[0] = \textrm{SoC}^0 C^s_i \label{eq:initial-conditions} \\[\smalleqnskip]
        & \qquad E^{\text{grid}}_m[t] = L_{m}[t] - \left(C^{\textrm{pv}} g_m^{\textrm{pv}}[t] + C^{\textrm{w}} g_m^{\textrm{w}}[t] - \nabla_m[t] \right) + \sum_i E_{i,m}[t] \label{eq:grid-variable} \\[\smalleqnskip]
        & \qquad \nabla_m[t] \leq C^{\textrm{pv}} g_m^{\textrm{pv}}[t] + C^{\textrm{w}} g_m^{\textrm{w}}[t] \label{eq:curtailment} \\[\smalleqnskip]
        & \qquad E^{\text{grid}}_m[t] \leq C^{\text{grid}} \Delta t \label{eq:grid-cap} \\[\smalleqnskip]
        & \qquad C^{\text{pv}} \leq C^{\text{pv}}_{\max} \\[\smalleqnskip]
        & \qquad p^{\textrm{pv}} C^{\textrm{pv}} + p^{\textrm{w}} C^{\textrm{w}} + \mathlarger{\sum}_i \: p^s_{i,m} C^s_i \leq B^{\text{cap}} \label{eq:budget-cap} \\[\smalleqnskip]
        % & \qquad C^s_i \leq z_i C^{\text{max}} \\[\smalleqnskip]
        % & \qquad \sum_i z_i = N \\[\eqnskip]
        %%
        \text{for all} & \qquad i \in [0,I{-}1], \: m \in [0,M{-}1], \: t \in [0,T{-}1] \tag*{}
        \end{align}
\end{subequations}

\begin{table}
    \centering
    \renewcommand{\arraystretch}{1.2}
    \begin{tabularx}{\linewidth}{ccX} \toprule \toprule
        Parameter & Units & \multicolumn{1}{>{\centering\arraybackslash}c}{Description} \\
        \midrule \midrule
        \multicolumn{3}{>{\centering\arraybackslash}l}{\small \quad Decision variables} \\
        $C^{\textrm{w}}$ & kWp & Installed offshore wind generation capacity \\
        $C^{\textrm{pv}}$ & kWp & Installed solar PV generation capacity \\
        $C^s_i$ & kWh & Installed energy capacity of storage technology $i$ \\
        $\nabla_m[t]$ & kWh & Generation curtailment at time $t$ in scenario $m$ \\
        $E_{i,m}[t]$ & kWh & Energy \textit{intake} to storage technology $i$ at time $t$ in scenario $m$ \\
        $\textrm{SoC}_{i,m}[t]$ & kWh & State-of-charge of storage technology $i$ at time $t$ in scenario $m$ \\
        \midrule
        \multicolumn{3}{>{\centering\arraybackslash}l}{\small \quad Derived variables} \\
        $E^{\text{grid}}_m[t]$ & kWh & Net energy \textit{drawn from} electricity grid at time $t$ in scenario $m$ \\
        $P^{\textrm{max}}_i$ & kW & Power capacity of storage technology $i$ \\
        \midrule
        \multicolumn{3}{>{\centering\arraybackslash}l}{\small \quad Sampled parameters} \\
        $\rho_m$ & -- & Probability of scenario $m$ \\
        $L_m[t]$ & kWh & Indsturial electrical load at time $t$ in scenario $m$ \\
        $g_m^{\text{w}}[t]$ & kWh/kWp & Wind power generation potential at time $t$ in scenario $m$ \\
        $g_m^{\text{pv}}[t]$ & kWh/kWp & Solar PV power generation potential at time $t$ in scenario $m$ \\
        $\eta_{i,m}$ & -- & Round-trip efficiency of storage technology $i$ in scenario $m$ \\
        $p^s_{i,m}$ & \euro/kWh/yr & Annualized capacity cost\textsuperscript{\textdagger} of storage technology $i$ in scenario $m$ \\
        \midrule
        \multicolumn{3}{>{\centering\arraybackslash}l}{\small \quad Known parameters} \\
        $\Delta t$ & hrs & Time step of simulation data \\
        $\nu_i$ & -- & Depth-of-discharge of storage technology $i$ \\
        $\delta_i$ & kW/kWh & Discharge ratio of storage technology $i$ (power capacity/energy capacity) \\
        $\textrm{SoC}^0$ & -- & Initial state-of-charge of storage (fraction of capacity) \\
        $p^{\text{w}}$ & \euro/kWp/yr & Annualized capacity cost\textsuperscript{\textdagger} of offshore wind \\
        $p^{\text{pv}}$ & \euro/kWp/yr & Annualized capacity cost\textsuperscript{\textdagger} of solar PV \\
        $p^e[t]$ & \euro/kWh & Price of grid electricity at time $t$ \\
        $p^c$ & \euro/kgCO$_2$ & Nominal carbon price \\
        $c[t]$ & kgCO$_2$/kWh & Carbon intensity of grid electricity at time $t$ \\
        $C^{\text{grid}}$ & kW & Grid connection capacity \\
        $C^{\text{pv}}_{\max}$ & kWp & Solar capacity limit \\
        $B^{\text{cap}}$ & \euro/yr & Annualized capital budget constraint \\
        \midrule
        \multicolumn{3}{>{\centering\arraybackslash}l}{\small \quad Indices} \\
        $i$ & -- & Storage technology \\
        $t$ & -- & Time step in modelled operation \\
        $m$ & -- & Scenario number \\
        \bottomrule \bottomrule
    \end{tabularx}
    \caption{Description of Stochastic Program variables \& parameters.} \label{tab:SP-params}
    \vspace{0.2cm}
    \footnotesize{\textsuperscript{\textdagger}\,Capital cost per energy/power capacity per year of lifetime}
\end{table}

\newpage

\subsubsection{Accounting for risk aversion} \label{sec:risk-averse-obj}

% Discuss CVaR tail up-weighting objective formulation, discuss motivation for this formulation (scale preservation which is important for comparability of VoI)
% Mention existing studies considering risk aversion \cite{mu2022CVaRbasedRiskAssessment,tostado-veliz2024RiskaverseElectrolyserSizing,pickering2019PracticalOptimisationDistrict} to motivate interest, however nobody yet considered risk-aversion when studying uncertainty reduction
% CVaR used to represent risk aversion in Linear Program \cite{rockafellar2000OptimizationConditionalValueatrisk}

The optimization objective in Eq. \ref{eq:SP-formualtion} assumes that the energy park designer is risk-neutral and is only interested in minimizing the average operating cost of the system, as is standard in the energy system design literature. However, when investing billions of Euros in large energy infrastructure projects, energy firms are typically risk-averse in their decision making, and \textit{are} concerned about the risk of high operating costs.
Previous studies have investigated the impact of risk aversion on energy system design, \edittwo{such as \cite{pickering2019PracticalOptimisationDistrict} \& \cite{tostado-veliz2024RiskaverseElectrolyserSizing}}, % also \cite{mu2022CVaRbasedRiskAssessment}
but none have considered how risk aversion affects the value of uncertainty reduction or optionality during decision making.

To account for the impact of risk aversion during energy park design, a risk measure is included in the objective function. The Conditional Value-at-Risk (CVaR) is used as it is a coherent risk measure compatible with Linear Programming \cite{rockafellar2000OptimizationConditionalValueatrisk}.
% and is what other studies use; (for thesis) maybe discuss choice of risk measures
For a linear scenario program where the cost and probability of each scenario are $c_m$ and $\rho_m$ respectively, an optimization minimizing the CVaR is given by,
\begin{subequations} \label{eq:cvar-defn}
    \begin{align}
        \addtocounter{equation}{-1}
        \min & \quad \xi + \frac{1}{\alpha} \sum\limits_{m} \rho_{m}\eta_{m}\\
        \text{s.t.} & \quad \eta_{m} \geq 0 \\
        & \quad \eta_{m} \geq c_{m} - \xi
    \end{align}
\end{subequations}

When optimized, this objective is equal to the expected value of the costs in the $\alpha$ right-tail of scenarios, $\mathbb{E}_{m:\, c_{m}>\xi_\alpha} \lbrace c_{m} \rbrace$, i.e. the average of the highest $\alpha\%$ of costs across the scenarios.\\
% For thesis: discuss how this happens and derive

As VoI and VoO are differences in average costs, to allow valid comparison between the risk-neutral and risk-averse cases, we develop a risk-averse objective which maintains the scale of the cost.
This objective is equivalent to a weighted average of the scenario costs, where the costs in the $\alpha\%$ right-tail are weighted by a factor $n{+}1$ relative to others. So, risk aversion is represented by the designer `caring' about high operating costs more than others, meaning the objective is still a physical cost, and so can be validly compared across cases.
This risk-averse objective is,
\begin{equation} \label{eq:risk-averse-obj}
    \frac{1}{1+n\alpha} \vast( \underbrace{\mathbb{E}\lbrace c_{m} \rbrace}_{\substack{\text{original}\\\text{obj.}}} + \ n\alpha \underbrace{\left(\xi + \frac{1}{\alpha} \sum\limits_{m} \rho_{m}\eta_{m} \right)}_{\text{CVaR}} \vast)
\end{equation}

where $c_m$ is the objective value in each scenario as specified in Eq. \ref{eq:SP-formualtion}, and both the original constraints for Eq. \ref{eq:SP-formualtion} and the constraints from Eq. \ref{eq:cvar-defn} are imposed.
When optimized this objective becomes $\left(\mathbb{E}\lbrace c_{m} \rbrace + n\alpha\, \mathbb{E}_{m:\, c_{m}>\xi_\alpha} \lbrace c_{m} \rbrace\right)/(1+n\alpha)$.
\section{Results \& Discussion} \label{sec:results}

% Aim for 2000 words

%\subsection{Prior design} \label{sec:prior-design}

% Discuss system design \& VoI for system sizing (compare solutions and costs, also mention benefit of adding storage to energy park)
%? Note optimized objective represents annualized cost of running system, assumed good enough

Initial designs for the energy park using each storage technology are produced by applying the Stochastic Program (SP) model (described in Sec. \ref{sec:SP}), using the prior distributions of renewable generation, industrial load, and storage technology performance (defined in Sec. \ref{sec:prob-gen-load} \& \ref{sec:prob-storage}).
Table \ref{tab:prior-comparison} compares the performance of these designs.
Sodium-Sulphur high-temperature batteries (NaS) are found to provide the lowest total system cost, so this design is selected, and NaS is chosen for procurement and R\&D. The design sizes \SI{441}{MW} wind and \SI{500}{MW} solar generation capacity, with \SI{1.51}{GWh} of NaS storage to support the system. The average total cost of this system is {\euro}124.1m/yr, with {\euro}10.7m/yr expected to be spent on batteries. If no energy storage were used in the energy park, \SI{486}{MW} of wind and \SI{500}{MW} of solar generation would be installed, leading to an average cost of {\euro}183.7m/yr. Therefore, installing bulk energy storage for arbitrage reduces the average cost of the energy park system by {\euro}59.6m/yr (32\%), and reduces the average operational carbon emissions by 20.9 ktCO$_2$/yr (20\%).\\

\begin{table}[h]
    \centering
    \renewcommand{\arraystretch}{1.25}
    \begin{tabular}{ccccc} \toprule \toprule
        Storage technology & \makecell{Total cost\\({\euro}m/yr)} & \makecell{Carbon emissions\\(ktCO$_2$/yr)} & \makecell{Storage capacity\\(GWh)} & \makecell{Cost of storage\\({\euro}m/yr)} \\
        \midrule \midrule
        None & 183.7 & 104.3 & -- & -- \\
        CAES & 148.1 & 84.2 & 3.59 & 7.2 \\
        Li-ion & 150.1 & 87.5 & 0.93 & 9.2 \\
        NaS & \textbf{124.1} & 83.4 & 1.51 & 10.7 \\
        VRFB & 176.6 & 93.9 & 0.40 & 5.2 \\
        \bottomrule \bottomrule
    \end{tabular}
    \caption{Performance of initial energy park designs using different storage technologies.}
    \label{tab:prior-comparison}
    \vspace{0.1cm}
    \footnotesize{\it Cost and carbon values are averages over prior distributions of load, generation, and storage performance.}
\end{table}

\subsection{\edit{Design updating without optionality \& the Value of Information}} \label{sec:design-wo-option}

\begin{figure}[h]
    \centering
    \includegraphics[width=0.7\linewidth]{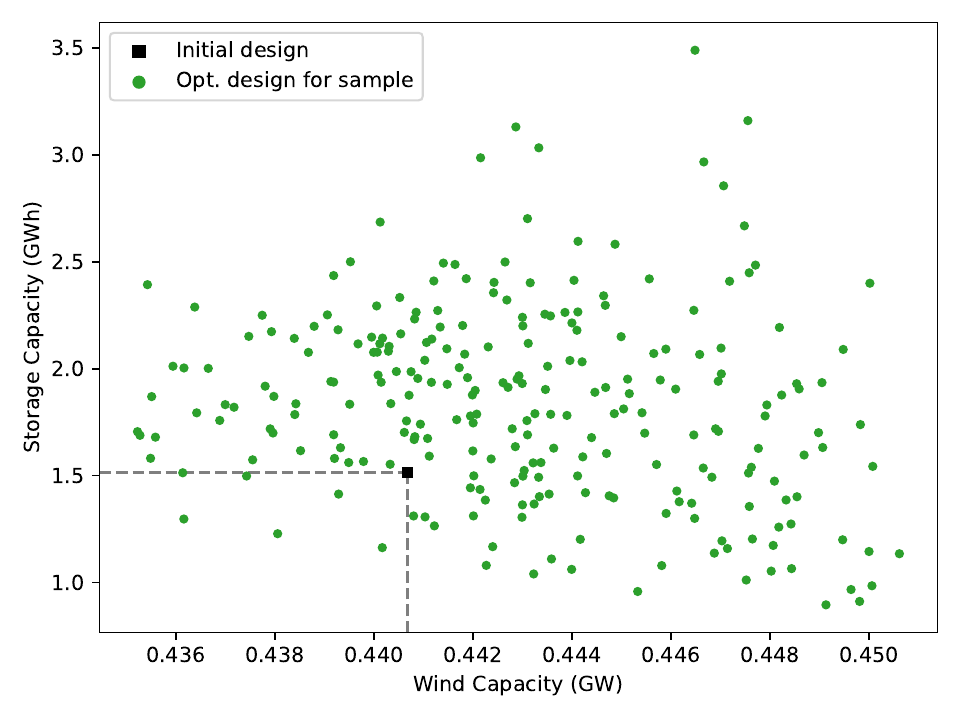}
    \caption{Final designs of energy park with NaS storage for each sampled measurement.}
    \label{fig:restricted-designs}
\end{figure}

To investigate the final design of the energy park given the uncertainty in the performance of NaS storage, 250 possible measurement values obtained from the demonstrator system built during R\&D were sampled for each performance parameter using the probabilistic model of storage performance (see Table \ref{tab:tech-prob-models}). % only mention z samples to improve clarity
For each sampled measurement, the SP model is applied using the corresponding posterior distribution of large-scale system performance to produce a final design.
Fig. \ref{fig:restricted-designs} plots the wind generation and NaS storage capacities of the final designs for each sampled measurement, and compares them to the initial design. All designs install the maximum \SI{500}{MW} of solar.

The average total cost of the final energy park design over the samples is {\euro}102.0m/yr, whereas the cost of the initial design was {\euro}124.1m/yr. Therefore, the Value of Information (VoI) associated with designing the energy park after R\&D has been performed to reduce uncertainty in the performance of NaS storage is {\euro}22.1m/yr (18\% of initial design cost). This cost reduction comes from the system designer's improved ability to trade off wind generation and storage capacity depending on the relative cost of energy arbitrage, and better make use of the capital budget. In some cases where the cost of NaS storage is low and its efficiency is high, the storage capacity in the final design is more than double that initially chosen. Assuming a 20 year project lifetime, having the option to \edit{update} the capacities of wind and storage procured from their suppliers after R\&D is worth {\euro}442m to the energy park developer. As a result, they should be willing to pay more (per unit capacity) to a supplier offering a flexible contract.\\

\subsection{\edit{Designing with optionality \& the Value of Optionality}} \label{sec:design-with-option}

% Discuss VoO - include lots of nice figures showing different tech selection \& performances
% Spend some time discussing interpretation and importance (discuss that convincing a company to do R\&D and demonstrator, and give `option to purchase agreement'/call option would be expensive - compare against VoO to determine if worthwhile, willingness to pay)

\begin{figure}
    \centering
    \includegraphics[width=0.65\linewidth]{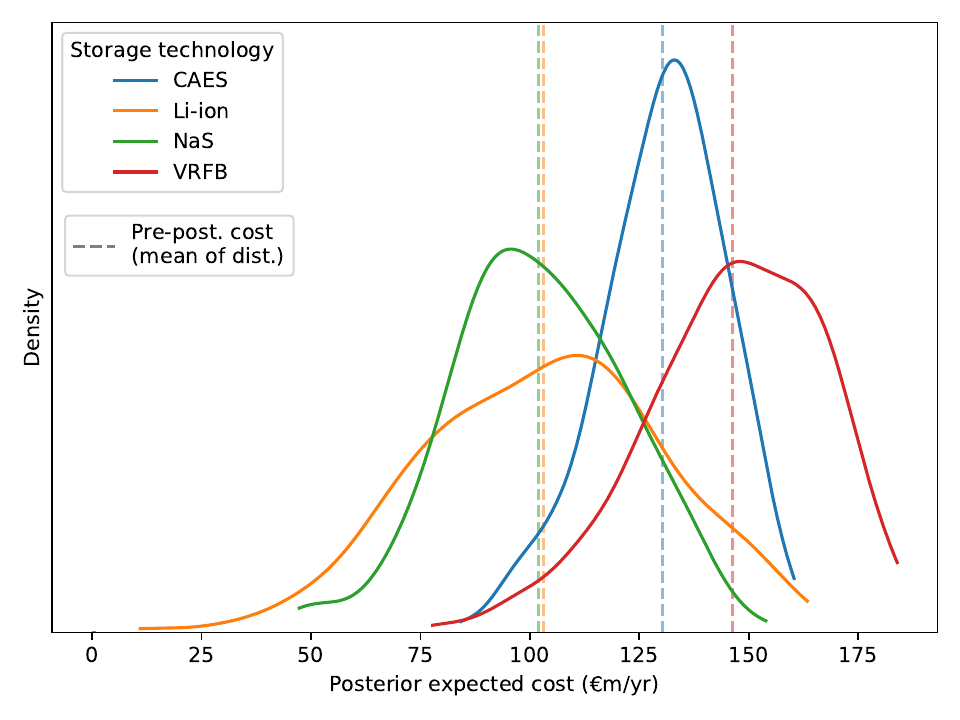}
    \caption{Distribution of final design expected costs over sampled measurements for each storage technology.}
    \label{fig:tech-cost-dists}
    \vspace*{0.1cm}
    \footnotesize{\it Dashed line indicates total cost of final designs averaged over sampled measurements, which is Pre-Posterior cost (Eq. \ref{eq:u-prepost}).}
\end{figure}

In the initial design, NaS was found to provide the lowest total cost \textit{on average}, and R\&D was performed only for this technology. However, it may be the case that NaS does not end up being the best technology, and after storage performance uncertainty is reduced another technology provides a lower average cost. Fig. \ref{fig:tech-cost-dists} plots the distribution of the expected costs of final designs for each storage technology, obtained by repeating the process of sampling measurement values and optimizing sizings (as described in the previous section). The significant overlap in the distributions indicates there are likely several cases where other storage technologies provide a lower average cost than NaS once performance uncertainty has been reduced.

The average cost of the energy park could therefore be reduced by performing R\&D with all storage technologies, allowing the best technology to be selected during final design when uncertainty has been reduced. However, performing R\&D, developing demonstrator systems, and purchasing supplier contracts  giving the right-to-purchase for all energy storage technologies comes at a significant cost.
% and ultimately only one technology is used
% so are the benefits of storage technology optionality worth the cost?

\begin{figure}
    \centering
    \includegraphics[width=0.8\linewidth]{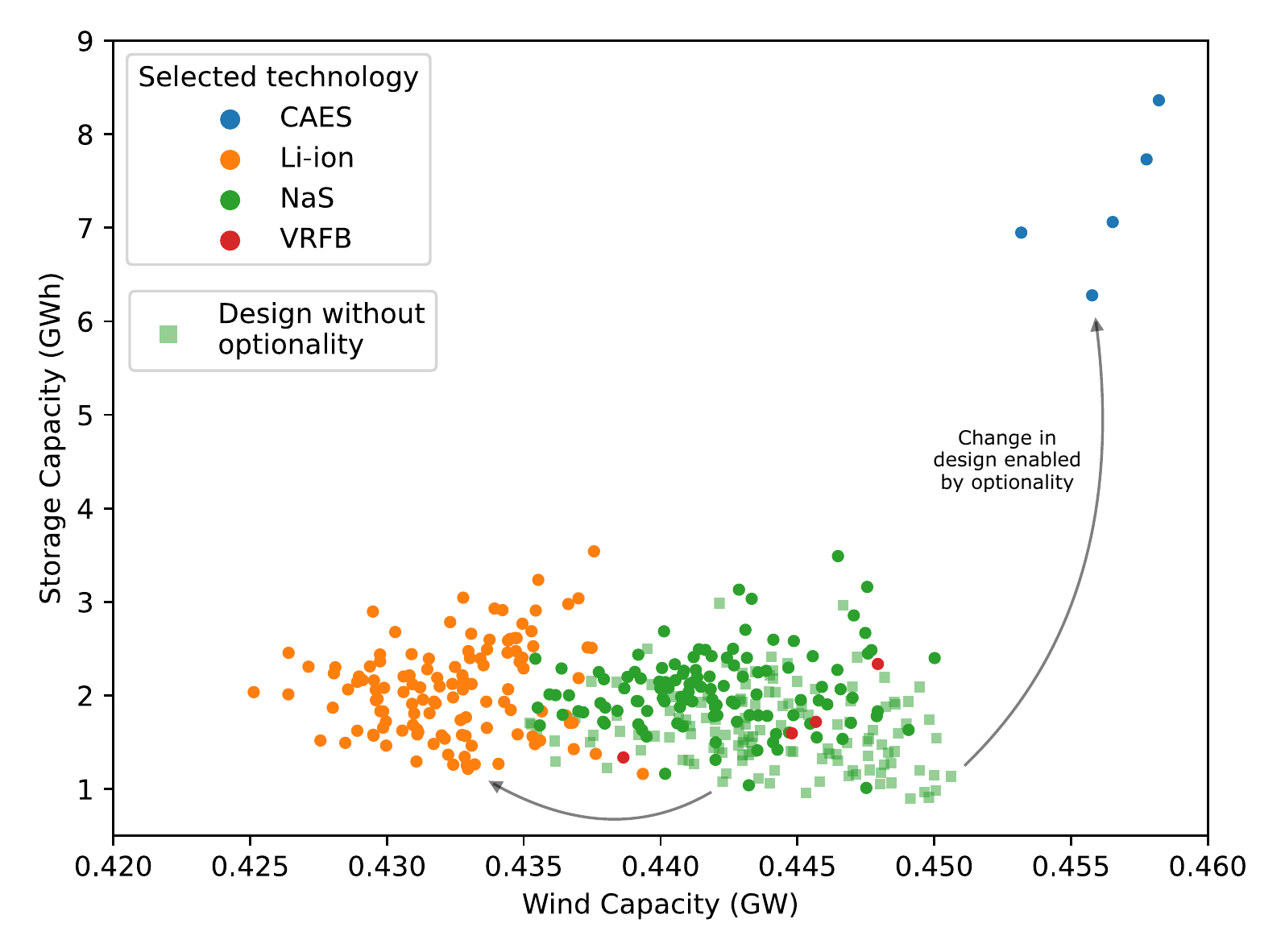}
    \caption{Final designs of energy park with storage optionality for each sampled measurement.}
    \label{fig:open-designs}
    \vspace*{0.1cm}
    \footnotesize{\it Marker color indicates the selected storage technology. Squares indicate final designs using NaS storage for samples where NaS does not provide the lowest average cost.}
\end{figure}

\begin{figure}
    \centering
    \includegraphics[width=\linewidth]{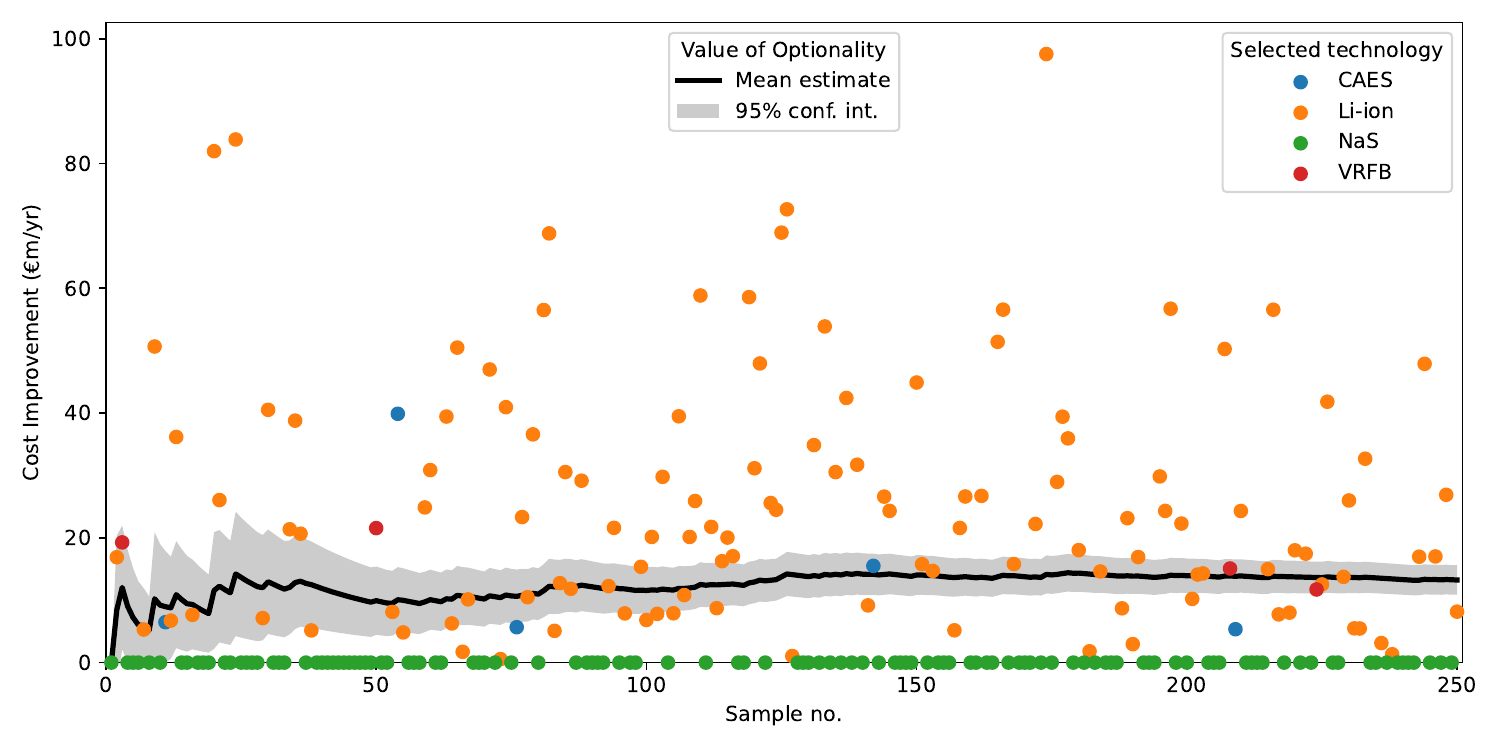}
    \caption{Reduction in average cost of final design provided by storage technology optionality, and selected technology, for each sampled measurement.}
    \label{fig:design-comparison}
\end{figure}

The process of sampling possible demonstrator measurement values is repeated (for all technologies), and final energy park designs are determined for each sample, this time allowing the best technology to be selected. Fig. \ref{fig:open-designs} plots the resulting final designs for each sample. The marker color indicates which storage technology is selected in each case. In 132 of 250 samples (53\%) a different technology than NaS is selected for the final design. In 123 of those cases Li-ion is the best technology choice after performance uncertainty has been reduced. Fig. \ref{fig:design-comparison} illustrates for each sample, how much the average cost of the energy park is reduced by being able to select the best storage technology, compared to the case where only NaS is available.

The average cost of the energy park designed \textit{with} storage technology optionality is {\euro}88.7m/yr, compared to {\euro}102.0m/yr if only NaS is available for final design. Therefore, the value of retaining storage optionality for the final design of the energy park (VoO) is {\euro}13.3m/yr (13\% of cost without optionality).
With a 20 year system lifetime, this optionality in storage technology selection is worth {\euro}266m to the energy park developer.
Long-duration energy storage demonstrator projects funded by the UK government have received around {\euro}1m for proof-of-concept systems \cite{desnz2023LongerDurationEnergy} % \cite{hydrostorinc2023HydrostorWrittenEvidence} - for thesis
and {\euro}10m for mid-scale demonstrators \cite{desnz2023LongerDurationEnergya}. % \cite{desnz2023LongerDurationEnergyb} - for thesis
So, the cost of developing and contracting Li-ion, VRFB, and CAES storage is conservatively estimated at {\euro}20m each.
Therefore, hedging bets on storage technology performance and maintaining optionality in technology choice through the energy park design process provides a net reduction in average cost of {\euro}206m (10\%) to the energy park developer.

\subsection{Sensitivity \edit{analysis}}

% Sensitivity analysis over $r$ parameter of measurement model - keep it brief, values change a bit, but not conclusion

A key assumption made in the probabilistic model of uncertain storage performance in Sec. \ref{sec:prob-storage} was the level of uncertainty reduction provided by measuring the demonstrator system \edit{built during R\&D}. This is encapsulated by the uncertainty reduction factor parameter $r$ in Eq. \ref{eq:measurement}, which quantifies how closely the \edit{designer believes the} performance of the demonstrator matches that of a large-scale system, and so how much uncertainty remains after measurement. Finding data from real-world storage systems to estimate this parameter is extremely challenging, as actual performance and cost data are highly commercially sensitive, and few large-scale systems currently exist.

Sensitivity analysis is performed to determine whether the conclusions provided by the Value of Optionality analysis remain valid. The process from Sec. \ref{sec:design-wo-option} \& \ref{sec:design-with-option} is repeated for values of $r$ over the feasible range. Table \ref{tab:r-sensitivity} shows VoI and VoO values for each case. For both VoI and VoO, the greater the uncertainty reduction provided by measurement (lower $r$), the greater the benefit to design. This is because lower uncertainty allows final designs with less hedging to be used, which achieve lower cost. While there will be some threshold of $r$ above which the VoO is less than the cost of R\&D, even with the lowest level of uncertainty reduction tested, $r=0.5$, retaining storage technology selection optionality is still economically worthwhile. So the conclusions of the base case analysis are robust to the level of uncertainty reduction. The variation of VoI with $r$ is larger than that of VoO, as VoO compares differences in designs both made with reduced uncertainty, so both costs in the difference benefit (reduce) from lower uncertainty.\\

\begin{table}[b]
    \centering
    \renewcommand{\arraystretch}{1.25}
    \begin{tabular}{ccc} \toprule \toprule
        Uncertainty reduction factor, $r$ & VoI ({\euro}m/yr) & Value of Optionality ({\euro}m/yr) \\
        \midrule \midrule
        0.1 & 26.4 & 15.1 \\
        0.2 & 23.5 & 13.9 \\
        \textbf{0.25} & 21.1 & 13.3 \\
        0.3 & 20.6 & 12.9 \\
        0.4 & 18.5 & 11.6 \\
        0.5 & 16.6 & 9.8 \\
        \bottomrule \bottomrule
    \end{tabular}
    \caption{Sensitivity of \edit{Value of Information} and Value of Optionality to level of uncertainty reduction.}
    \label{tab:r-sensitivity}
\end{table}
% Julia (Stanford) had a really nice comment about this; Building the demonstrator takes time and potentially delays the project. But the more time and effort you spend on it, potentially the better information you get out of it (lower uncertianty). So if you could make a correspondence between the delay in building and better demonstrator and the associated uncertainty reduction r, you could compare against the cost of that delay to work out the best amount of time and effort to spend building the demonstrator. Nice!
% Another comment from the STEER seminar was; As demonstrators are built, the uncertainty in storage performance will change not only for this project, but also for all future projects, which means there is additional value to building demonstrators, and the method needs to be rerun for each project with the current remaining level of uncertainty (updated prior), and over time the benefit of optionality will reduce because the uncertainty in storage performance is decreasing.

\edit{During operation the energy park is able to trade with the grid to reduce the cost of the energy it supplies to the hydrogen plant, but must pay a cost for any associated carbon emissions. However, due to lack of data availability, uncertainty in grid electricity price and carbon emissions were not included in the uncertainty model. So, a similar sensitivity analysis is performed to check whether the pricing and carbon data assumed in the base case are representative, or whether changes in this data have a significant effect on the resulting VoI and VoO values.}

\edit{Three alternative cases are tested: the 2023 pricing data is increased and decreased by 10\% respectively, and pricing and carbon data from 2024 is used. Table \ref{tab:price-sensitivity} presents the VoI and VoO values for each of these cases.}
\edit{While the VoI is found to vary significantly, by up to 75\%, in all of the alternative cases it is greater than the base case. This suggests that design flexibility could be even more valuable than indicated by the initial results.
As before, the changes in VoO values are smaller than for the VoI. And again, in all cases the VoO is greater than the cost of developing the additional storage technologies, so retaining optionality is still found to be economically worthwhile.
Therefore, the conclusions from Sec. \ref{sec:design-wo-option} \& \ref{sec:design-with-option} are not affected by the electricity price and carbon emissions data used.}\\

\begin{table}[t]
    \centering
    \renewcommand{\arraystretch}{1.33}
    \begin{tabular}{cccc} \toprule \toprule
        Case & \makecell{Prior operating cost\\({\euro}m/yr)} & \makecell{Value of Information\\({\euro}m/yr)} & \makecell{Value of Optionality\\({\euro}m/yr)} \\
        \midrule \midrule
        \makecell{2023 prices \& carbon\\(base case)} & 124.1 & 22.1 & 13.3 \\
        2023 prices $+10\%$ & 124.9 & 39.2 & 16.1 \\
        2023 prices $-10\%$ & 151.5 & 32.1 & 12.1 \\
        2024 prices \& carbon & 80.9 & 38.2 & 17.1  \\
        \bottomrule \bottomrule
    \end{tabular}
    \caption{\edit{Sensitivity of Value of Information and Value of Optionality to electricity prices and carbon emissions}}
    \label{tab:price-sensitivity}
\end{table}

\edit{Whether or not retaining optionality in storage technology selection is worthwhile is determined by the comparison between the VoO and the cost of retaining optionality. The cost of optionality is therefore a critical driver of this conclusion.
Were the \euro20m cost of building a demonstrator system used in the initial analysis (see Sec. \ref{sec:design-with-option}) to double, this would correspond to an annualized cost of {\euro}6m/yr for developing all three additional storage technologies. However, even this doubling of a pessimistic cost estimate is still significantly smaller than any of the VoO values found across the sensitivity analyses, the smallest of which is \euro9.8m/yr from the extreme case of low uncertainty reduction, $r=0.5$.
So, the assumptions made about the cost of optionality do not affect the conclusions drawn from the analysis, as the VoO is sufficiently large to provide strong confidence that storage optionality is worthwhile for the design of the studied energy park.}

\subsection{Use of multiple storage technologies}\label{sec:two-techs}

% Results for energy park with two storage technologies
% Discuss importance for energy park design - two technologies is beneficial, lowers costs (could mention different behaviours of each technology); provide both reduced cost and reduced VoO values (implies using two storage technologies greatly improvements design robustness to uncertainty in storage performance)

So far it has been assumed that only a single energy storage technology is installed in the energy park to perform arbitrage. The different characteristics of the energy storage technologies (such as efficiency, cost per energy capacity, and discharge ratio) make them suited for arbitrage over different time scales \cite{irena2017ElectricityStorageRenewables}. As trends in renewable generation and grid electricity price also occur over a range of time scales, the profit from energy arbitrage could be increased by installing multiple energy storage technologies, and sizing them for the volume of the arbitrage they are best suited to.

\begin{figure}[t]
    \centering
    \includegraphics[width=\textwidth]{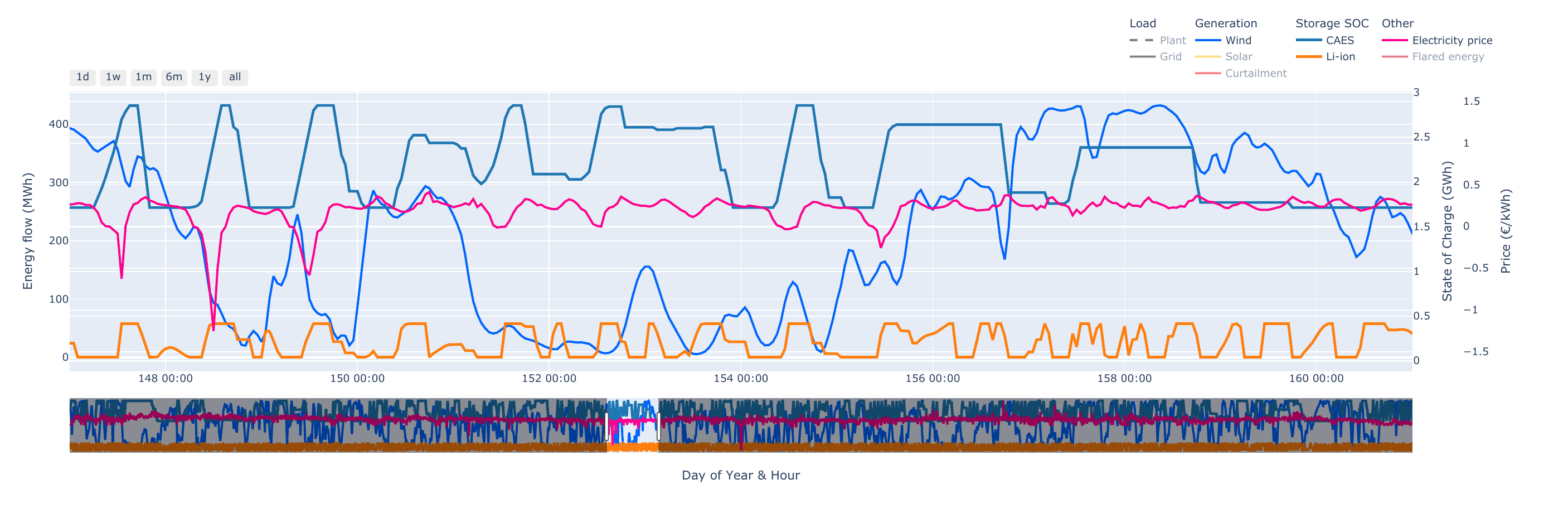}
    \caption{Operation of energy park with Li-ion and CAES storage, demonstrating different energy arbitrage time scales.}
    \label{fig:op-sim}
    \vspace{0.1cm}
    \footnotesize{\textit{Interactive version of plot available online} \href{https://mal84emma.github.io/Energy-Park-Design/CAES-Li-ion_operation_plot.html}{\textbf{here}}.}
\end{figure}

The investigation is repeated for the case where two storage technologies are used in the energy park. This means two technologies are selected and developed following the initial design. Fig. \ref{fig:op-sim} plots an example of the operation of the energy park with two storage technologies, for the initial system design with Li-ion and CAES. The state-of-charge traces for Li-ion (orange) and CAES (dark blue) demonstrate the different arbitrage behaviours of the two technologies.
\edit{As the Linear Program simulating the energy park (Eq. \ref{eq:SP-formualtion}) simultaneously optimizes over all operational variables, it is able to dynamically dispatch the two storage technologies, scheduling them to best suit their characteristics and maximize the combined performance. In this case, best performance is achieved by operating the} Li-ion at short time scales with small traded energy volumes, while the CAES \edit{is operated} over longer time scales with greater energy volumes.

% Prior cost - 113.97 €m/yr, ['Li-ion', 'NaS']
% Restricted expected cost - 96.08 €m/yr
% Open expected cost - 94.46 €m/yr
% VoI - 17.88 €m/yr (15.7\%)
% VoO - 1.62 €m/yr (1.7\%)
% And much more even distribution of tech combo selections

\begin{table}[!b]
    \centering
    \renewcommand{\arraystretch}{1.25}
    \begin{tabular}{cccccc} \toprule \toprule
        & Design case & \makecell{Total cost\\({\euro}m/yr)} & \makecell{VoI/{\color{gray!50!black}VoO}\\({\euro}m/yr)} & \makecell{Carbon emissions\\(ktCO$_2$/yr)} & \makecell{Cost of storage\\({\euro}m/yr)} \\
        \midrule \midrule
        & No storage & 183.7 & -- & 104.3 & -- \\
        \arrayrulecolor{black!50}\midrule
        \parbox[t]{2mm}{\multirow{3}{*}{\rotatebox[origin=c]{90}{1 tech.}}}
        & Initial design & 124.1 & -- & 83.4 & 10.7 \\
        & Design with flexibility* & 102.0 & 22.1 & 79.8 & 13.0 \\
        & Design with optionality & 88.7 & {\color{gray!50!black}13.3} & 79.2 & 14.4 \\
        \arrayrulecolor{black!50}\midrule
        \parbox[t]{2mm}{\multirow{3}{*}{\rotatebox[origin=c]{90}{2 techs.}}}
        & Initial design & 114.0 & -- & 80.6 & 11.2 \\
        & Design with flexibility* & 87.5 & 26.5 & 78.9 & 15.1 \\
        & Design with optionality & 85.7 & {\color{gray!50!black}1.8} & 79.0 & 15.0 \\
        \bottomrule \bottomrule
    \end{tabular}
    \caption{Comparison of energy park performance for different design cases.}
    \label{tab:two-techs-comparison}
    \vspace{0.1cm}
    \footnotesize{\it All values are averages over the relevant distributions. *Updated (final) designs without optionality.}
\end{table}

For the energy park with two technologies, the average cost of the final designs is {\euro}87.5m/yr for the case without optionality, and {\euro}85.7m/yr for the case with optionality.
Table \ref{tab:two-techs-comparison} compares these costs to the results for the system with only one storage technology from Sec. \ref{sec:design-wo-option} \& \ref{sec:design-with-option}. While using two storage technologies provides a substantial cost reduction for the case without optionality in the final design, {\euro}14.5m/yr (14\%), with optionality this cost reduction falls to {\euro}3.0m/yr (3\%). However, this cost saving is still significant ({\euro}60m over a 20yr system lifetime), and shows that multiple energy storage technologies should be used for arbitrage in energy parks.

The VoO is now {\euro}1.8m/yr (2.1\%), or {\euro}36m over the 20 year lifetime. This is less than the estimated {\euro}40m cost of developing the remaining two storage technologies not selected in the initial design. So, when two technologies are used, it is not worthwhile hedging bets against uncertain storage performance by retaining optionality. This means selecting the best two storage technologies is robust to the performance uncertainty. The VoI however remains high, {\euro}26.4m/yr, meaning that contract flexibility and updating the final system design are still very valuable.

%\newpage
\subsection{Risk-averse design}

% Results with risk aversion in objective, Eq. \ref{eq:risk-averse-obj}
% Looking at Fig. \ref{fig:design-comparison}, for some samples the VoO is very high, meaning the energy park developer could lose out a lot (motivates risk aversion)
% Briefly discuss impact of risk aversion on prior costs vs pre-posterior costs - i.e. impact on VoI vs VoO
% Suggest little impact on VoO as tail risks similar for all storage technologies (uncertainties have same shape)

% No storage cost: 195.91 €m/yr
% Prior cost: 136.55 €m/yr, ['NaS']
% Restricted expected cost: 114.76 €m/yr
% Open expected cost: 101.46 €m/yr
% VoI: 21.79 €m/yr (16.0%)
% VoO: 13.30 €m/yr (11.6%)
% costs are higher, indicating risk-aversion is working

The objective function used to design the energy park in previous sections, Eq. \ref{eq:SP-formualtion}, assumes that the designer is risk-neutral and is only interested in the average system cost. However, for capital-intensive energy infrastructure projects at this scale, energy firms are usually risk-averse and concerned by the possibility of poor financial returns. The effect of risk aversion on the energy park design, and the benefits of uncertainty reduction (VoI) and optionality (VoO) for improving design, are studied by repeating the investigation using the risk-averse objective defined in Eq. \ref{eq:risk-averse-obj}, which increases the weighting on the worst $\alpha\%$ of scenarios $n$-fold in the cost average.

For the case where $\alpha=10\%$ and $n=1$, adding risk aversion increases the expected cost of the initial design from \euro124.1m/yr in the risk-neutral base case to \euro136.6m/yr, with NaS again being selected as the best storage technology. This cost increase occurs as the system designer is now more pessimistic, and puts more weighting on the highest 10\% of costs across the scenarios. However, whilst the expected cost increases by 10\%, the system sizing remains unchanged. The same pattern is found for the final designs, with costs increasing by 12.5\% on average but the designs changing negligibly.

Table \ref{tab:risk-averse-results} presents the VoI and VoO values calculated for various combinations of $\alpha$ and $n$. Increasing risk aversion causes a small decrease in VoI, whereas the impact on VoO is negligible. This indicates that the cost variability for the final designs, and so the risk penalty, is the same across the storage technologies. So this variability must be driven by the uncertainties in renewable generation and industrial load (see Table \ref{tab:ts-prob-models}) rather than the storage uncertainties. % could do some analysis to demonstrate this?; note that generation and load uncertainties not reducable?
Therefore, while reducing storage uncertainty is important for system design, it does not have a significant impact on risk.\\
% There is significant risk associated with the cost of the energy park, which is important to consider for the financial planning of the project. However little can be done to reduce this risk as the system design does not significantly change with risk aversion, and the uncertainties in generation and load cannot be reduced at the time of system design. % accurate models of generation and load really important

\begin{table}[h]
    \centering
    \renewcommand{\arraystretch}{1.25}
    \begin{tabular}{cccc} \toprule \toprule
        Confidence level, $\alpha$ (\%) & Tail-weighting, $n$ & VoI ({\euro}m/yr) & VoO ({\euro}m/yr) \\
        \midrule \midrule
        \textbf{10} & \textbf{1} & 21.8 & 13.3 \\
        & 2 & 21.5 & 13.4 \\
        & 5 & 20.8 & 13.5 \\
        25 & 1 & 22.1 & 13.4 \\
        & 2 & 22.0 & 13.5 \\
        \midrule
        \multicolumn{2}{>{\centering\arraybackslash}c}{\small Base case (no risk aversion)} & 21.1 & 13.3 \\
        \bottomrule \bottomrule
    \end{tabular}
    \caption{Value of Information and Value of Optionality for risk-averse design cases}
    \label{tab:risk-averse-results}
\end{table}

% \subsection{Practical importance \& limitations of study}
% Include if space
% Why are these results important? What are the limitations?
\newpage
\vspace*{0cm}
\section{Conclusions} \label{sec:conclusions}

% Aim for c. 500 words
% Summarise results and importance.

This study investigated the impact of uncertainty in the cost, lifetime, and efficiency of energy storage technologies on energy park design, and the benefit of retaining optionality in storage technology choice through the design process. It quantified the improvement in design (reduction in average total cost) achieved when multiple technologies are developed to retain optionality \edit{in technology choice,} and allow a better storage \edit{selection} to be made \edit{after uncertainty in performance has been reduced}.
An illustrative energy park system containing wind and solar generation and bulk energy storage to support a hydrogen electrolyser, modelled off a real-world proposal in the Port of Rotterdam, was used for experiments. Four candidate bulk energy storage technologies were considered: Lithium-ion batteries (Li-ion), Sodium-Sulphur high-temperature batteries (NaS), Vanadium redox-flow batteries (VRFB), and compressed-air energy storage (CAES).

For the base case where only a single storage technology was selected for development, updating the system design (wind, solar, and storage sizings) after R\&D, when an improved estimate of storage performance is available (i.e. with reduced uncertainty), was found to reduce total costs by 18\% on average (VoI of \euro442m). This demonstrates that flexibility in procurement contracts for generation and storage is highly valuable to energy park developers.

If all four energy storage technologies were developed, so that when the system design was \edit{updated} any technology could be selected for the final design, the average total cost was found to reduce by 13\% compared to the case without optionality (VoO of \euro266m). As the cost of \edit{performing} R\&D for three additional storage technologies was estimated at \euro60m, retaining optionality in storage technology selection through the design process is economically worthwhile, and reduces the net cost of the energy park project by \euro206m on average.

Installing two storage technologies in the energy park to perform energy arbitrage at different time scales reduced the total system cost by 14\% on average (\euro290m) compared to the base case without optionality. And in this, case developing the two extra storage technologies to provide optionality was determined to not be worthwhile, reducing average costs by \euro36m (VoO) compared to the \euro40m additional cost of R\&D. The VoI however increased to \euro528m, so updating the system design is more valuable when trade-offs can be made between \textit{two} storage technologies and generation.

The decision-making recommendations from the VoI and VoO results were shown to be robust to the level of uncertainty reduction in energy storage performance provided by R\&D used in the statistical model, \edit{as well as the grid electricity price and carbon emissions data used, and the cost of developing demonstrator storage systems assumed.}
Additionally, including risk aversion in the design objective was found to have little impact on the optimized system designs or VoI and VoO results, as the risk (variability in system cost) was driven mostly by uncertainty in renewable generation and industrial load.

These results provide two important insights for energy park developers. Firstly, \edit{updating} system designs after improved estimates of storage technology performance have been gathered (i.e. uncertainty reduced) provides substantial savings in system cost. Hence, flexible procurement contracts should be negotiated to allow design adjustments to be made as better information becomes available. Secondly, two energy storage technologies should be used to support the energy park. Doing so both significantly reduces the \edit{total} cost of the system, % say why?
and means that a robust choice of storage technologies can be made up-front, as hedging against the chance of a different technology providing better performance is not worthwhile if two technologies are developed and installed.

\edit{A significant limitation of the VoI methodology is that it cannot provide generalization guarantees for the decision support recommendations it provides. While this study demonstrates that for the design of the particular energy park considered, design flexibility and storage optionality provide significant value, further work is required to determine whether these insights apply to other energy park systems. For example, those with different industrial loads, different generation mixes, or in geographic locations with different weather patterns and so renewable generation uncertainties.
However, this work does demonstrate the importance of performing this type of uncertainty analysis using the VoI framework when designing energy parks. As it shows that the total cost of an energy park system can be substantially reduced (in this case by 30\%) by enabling the system design to be updated with optionality in storage technology choice after uncertainty in the performance of those storage technologies has been reduced.}

\edit{The methodology should also be used to study the importance of reducing uncertainty and providing optionality in other components of an energy park. For instance renewable generation technologies, where uncertainty in their energy production is driven by technical parameters and local site characteristics, which could be reduced by equipment testing and site surveys, that may then influence the site selection decision.}
\newpage
\section*{CRediT authorship contribution statement}

\textbf{Max Langtry}: Conceptualization, Software, Methodology, Investigation, Writing - Original Draft, Writing - Review \& Editing
\textbf{Ruchi Choudhary}: Supervision, Writing - Review \& Editing

\section*{Declaration of competing interests}

The authors declare that they have no known competing financial interests or personal relationships that could have appeared to influence the work reported in this paper.

\section*{Data availability}

All code and data used to perform the experiments in this study are available at \url{https://github.com/mal84emma/Energy-Park-Design}. The numerical results presented are available at \url{https://zenodo.org/records/15050619}.

\section*{Acknowledgements}

Max Langtry is supported by the Engineering and Physical Sciences Research Council, through the CDT in Future Infrastructure and Built Environment: Resilience in a Changing World, Grant [EP/S02302X/1].

This work was performed using resources provided by the Cambridge Service for Data Driven Discovery (CSD3) operated by the University of Cambridge Research Computing Service (\href{https://www.csd3.cam.ac.uk/}{www.csd3.cam.ac.uk}), provided by Dell EMC and Intel using Tier-2 funding from the Engineering and Physical Sciences Research Council, Grant [EP/T022159/1], and DiRAC funding from the Science and Technology Facilities Council (\href{https://dirac.ac.uk/}{www.dirac.ac.uk}).

\newpage
\appendix
\section{Parameter settings used in experiments} \label{app:case-study-params}

% Maybe could delete text and just have tables with detailed captions
% Common parameter values specifying the district energy system, probabilistic model of building load, and procedure for sampling from that model, are used across all experiments. Table \ref{tab:system-params} provides values for the parameters of the district energy system model, including costs defining the objective/utility, simulation durations, and assumptions regarding battery performance and grid control. All parameters for the probabilistic model of building load are specified within the definition of the probabilistic model in Sec. \ref{sec:prob}. Table \ref{tab:sampling-params} provides the settings used when sampling for the defined distributions, and performing scenario reduction using the Fast Forward technique \cite{heitsch2003ScenarioReductionAlgorithms}, as implemented in \cite{gioia2023ScenarioReducer}.\\

% system parameters
\begin{table}[h]
    \centering
    \renewcommand{\arraystretch}{1}
    \renewcommand\cellset{\renewcommand\arraystretch{0.5}%
        \setlength\extrarowheight{0pt}}
    \begin{tabularx}{\linewidth}{lccX} \toprule \toprule
        \multicolumn{1}{>{\centering\arraybackslash}c}{Parameter} & Units & Value & \multicolumn{1}{>{\centering\arraybackslash}c}{Note/Refs} \\
        \midrule \midrule
        Simulation duration, $T$ & Hours & 8760 & \\
        Simulation time step, $\Delta t$ & Hours & 1 & \\
        Initial state-of-charge, $\text{SoC}^0$ & -- & 0.75 & \\

        Offshore wind CAPEX & \euro/kWp & 5000 & \cite{stehly20232022CostWind}, used to derive annualized capacity cost, $p^\text{w}$ \\
        Offshore wind OPEX & \euro/kWp/yr & 100 & \texttt{"} \\
        Offshore wind lifetime & Years & 20 & \texttt{"} \\
        Solar PV CAPEX & \euro/kWp & 1000 & \cite{ramasamy2023SolarPhotovoltaicSystem}, used to derive annualized capacity cost, $p^\text{pv}$ \\
        Solar PV OPEX & \euro/kWp/yr & 10 & \texttt{"} \\
        Solar PV lifetime & Years & 20 & \texttt{"} \\
        
        Carbon cost, $p^c$ & \euro/kgCO$_2$ & 1 & \makecell[tl]{Chosen to be significantly larger than current\\carbon trading prices, c. \euro0.1/kgCO$_2$ \cite{desnz2022UKETSCarbon,bloomberg2024EUETSMarket},\\to reflect incentive to produce green hydrogen.} \\
        Grid capacity, $C^{\text{grid}}$ & MW & 500 & \\
        Solar capacity limit, $C^{\text{pv}}_{\max}$ & MW & 500 & \\
        Capital budget & {\euro}m/yr & 200 & Corresponds to {\euro}4bn over a 20 year lifetime \cite{auchincloss20242023YearDelivery} \\
        \midrule
        \multicolumn{4}{>{\centering\arraybackslash}l}{\small\it Storage parameters} \\
        Depth-of-discharge, $\nu_i$ & \% & 90 & Li-ion, taken from \cite{irena2017ElectricityStorageRenewables} \\
        & & 100 & NaS, \texttt{"} \\
        & & 100 & VRFB, \texttt{"} \\
        & & 40 & CAES, \texttt{"} \\
        Discharge ratio, $\delta_i$ & kWp/kWh & 2 & Li-ion, taken from \cite{kebede2022ComprehensiveReviewStationary} \\
        & & 1 & NaS, \texttt{"} \\
        & & 0.5 & VRFB, \texttt{"} \\
        & & 0.1 & CAES, \texttt{"} \\
        \bottomrule \bottomrule
    \end{tabularx}
    \caption{Parameter values for energy system model used in experiments.}
    \label{tab:system-params}
\end{table}

\hfill \\

% sampling parameters
\begin{table}[h]
    \centering
    \renewcommand{\arraystretch}{1}
    \begin{tabular}{lc} \toprule \toprule
        \multicolumn{1}{>{\centering\arraybackslash}c}{Parameter} & Value \\
        \midrule \midrule
        No. samples from prior distribution & 250 \\
        No. samples from each posterior distribution & 250 \\
        MCMC sampling burn-in period & 250 \\
        MCMC sampling thinning factor & 10 \\
        No. of reduced scenarios used in Stochastic Program & 25 \\
        \bottomrule \bottomrule
    \end{tabular}
    \caption{Settings for sampling from probabilistic models used in experiments.}
    \label{tab:sampling-params}
\end{table}

\newpage
\bibliographystyle{elsarticle-num} 
\bibliography{EP-VOI_refs}

\end{document}